\documentclass[11pt]{article}
\usepackage[a4paper,margin=20mm]{geometry}
\usepackage{amsmath,amssymb,amsthm}
\usepackage{microtype,hyperref,titling}

\newtheorem{lemma}{Lemma}
\newtheorem{corollary}{Corollary}
\newtheorem{proposition}{Proposition}
\newtheorem{remark}{Remark}
\newtheorem{example}{Example}

\title{Relative Entropy under Ordered Chi-Squared Constraints: Sharp Bounds and Extremizers}

\author{Igal Sason\\[0.4em]
  \small Viterbi Faculty of Electrical and Computer Engineering\\
  \small and the Faculty of Mathematics\\
  \small Technion--Israel Institute of Technology, Haifa 3200003, Israel\\
  \small E-mail: \texttt{eeigal@technion.ac.il}}

\date{\today}

\begin{document}

\maketitle

\begin{abstract}
We derive a sharp upper bound on the relative entropy $D(P\|Q)$ in
terms of the two ordered chi-squared divergences $\chi^2(P\|Q)$ and
$\chi^2(Q\|P)$. For any $A,B>0$, the constraints $\chi^2(P\|Q)\leq A$
and $\chi^2(Q\|P)\leq B$ yield upper bounds on $D(P\|Q)$ and
$D(Q\|P)$. Two pairs of Bernoulli distributions each attain both
bounds simultaneously. As a consequence, for sequences of probability
mass functions $\{P_n\}$ and $\{Q_n\}$ on a common finite or countably infinite
alphabet, both $D(P_n\|Q_n)$ and $D(Q_n\|P_n)$ tend to zero whenever
one of the two ordered chi-squared divergences tends to zero and the
other remains uniformly bounded. Without this boundedness assumption,
$D(P_n\|Q_n)$ can fail to converge to zero even when
$\chi^2(Q_n\|P_n)\to 0$. Finally, in contrast to the attainable upper
bounds, on any finite or countably infinite alphabet containing at
least three elements, the infimum of $D(P\|Q)$ under the exact
constraints $\chi^2(P\|Q)=A$ and $\chi^2(Q\|P)=B$ is zero and is
not attained. As an application to lossless source coding, we consider the product
measures $P^n$ and $Q^n$ induced by $P$ and $Q$, respectively, on
blocks of length $n$. We bound the expected saving in ideal codelength
under $P^n$ when the encoder replaces $Q^n$ with the mixture
$\lambda P^n+(1-\lambda)Q^n$. The lower bound depends on the sources
only through $D(P\|Q)$, and the upper bound depends on them only
through the two ordered chi-squared divergences between $P$ and $Q$.
For every fixed $n$, the upper bound agrees with the expected saving
to first order as $\lambda \to 0^+$.
\end{abstract}



\section{Introduction}
\label{section: introduction}

How large can the penalty for distribution mismatch be if, under its respective reference 
distribution, each likelihood ratio has a second moment bounded by a prescribed constant?

Relative entropy quantifies this penalty in both lossless source
compression and statistical prediction. 
Suppose that a source emits a symbol $X$ according to a probability
mass function $P$, but a lossless source code is designed using a
probability mass function $Q$. With ideal code lengths, the excess
expected description length due to this mismatch is
\[
 \mathbb E_P[-\log Q(X)]-\mathbb E_P[-\log P(X)]
 = D(P\|Q),
\]
provided that both expectations are finite. Throughout, $\log$ denotes
the logarithm to a fixed base $c>1$. When $c=2$, this penalty
is measured in bits per source symbol. In statistical prediction, a
forecaster who assigns probability $Q(x)$ to outcome $x$ incurs
logarithmic loss $-\log Q(x)$ when $x$ occurs. The same identity
therefore gives the excess expected logarithmic loss incurred by
using $Q$ in place of $P$.

To formulate the problem precisely, let $P$ and $Q$ be probability
mass functions on a common finite or countably infinite alphabet
$\mathcal X$. We use the discrete setting to simplify the notation.
The results also hold for probability measures on a general
measurable space. We write
\begin{align}
 D(P\|Q)&:=\sum_{x\in\mathcal X} P(x)\log\frac{P(x)}{Q(x)},\\
 A&:=\chi^2(P\|Q)
   =\sum_{x\in\mathcal X}\frac{(P(x)-Q(x))^2}{Q(x)},\\
 B&:=\chi^2(Q\|P)
   =\sum_{x\in\mathcal X}\frac{(P(x)-Q(x))^2}{P(x)}.
\end{align}
We use the conventions $0\log(0/q)=0$ for $q\geq 0$ and $0/0=0$.
A positive mass at a point of zero reference mass contributes
$+\infty$ to the corresponding divergence. Thus, if both $A$ and
$B$ are finite, then $P$ and $Q$ have the same support. As an
expected log-likelihood ratio, $D(P\|Q)$ remains well defined even
when both expected logarithmic losses are infinite and their
difference is undefined.

When $P$ and $Q$ have the same support, the second moments of the
two likelihood ratios satisfy
\[
 \mathbb E_Q\left[\left(\frac{P(X)}{Q(X)}\right)^2\right]=1+A,
 \qquad
 \mathbb E_P\left[\left(\frac{Q(X)}{P(X)}\right)^2\right]=1+B.
\]
Thus, upper bounds on these moments are equivalent to upper bounds
on the two ordered chi-squared divergences. Given $a,b\geq 0$, we
seek the largest possible mismatch penalty $D(P\|Q)$ among all pairs
$(P,Q)$ satisfying
\[
 \chi^2(P\|Q)\leq a,\qquad \chi^2(Q\|P)\leq b.
\]

The same three divergences arise in multinomial
goodness-of-fit testing. If $P$ is the empirical distribution of
$n$ observations and $Q$ is the null distribution, then
$n\chi^2(P\|Q)$ is Pearson's statistic, $n\chi^2(Q\|P)$ is
Neyman's modified chi-squared statistic, and $2nD(P\|Q)/\log e$
is the likelihood-ratio statistic. Basu and
Sarkar~\cite{BasuSarkar1994} study multinomial goodness-of-fit
tests based on disparities, including the power-divergence
family to which these statistics belong.

The classical inequality
\begin{equation}
\label{eq: well-known}
 D(P\|Q)\leq \log\bigl(1+\chi^2(P\|Q)\bigr)
\end{equation}
uses only the chi-squared divergence in the forward order and
follows directly from Jensen's inequality, or from the monotonicity
of the R\'{e}nyi divergence in its order~\cite{ErvenH14}.
The weaker bound \[ D(P\|Q)\leq (\log e)\,\chi^2(P\|Q) \] appears,
for example, in a paper by Csisz\'{a}r and
Talata~\cite[Lemma~6.3]{CsiszarTalata2006} on context-tree estimation,
which concerns learning from data the variable-length memory
structure of a stochastic process. The reverse chi-squared divergence
carries additional information even on a binary alphabet. In
particular, if the forward chi-squared divergence is held fixed
and the reverse chi-squared divergence tends to zero, must
$D(P\|Q)$ also tend to zero? We answer this question in the
affirmative by determining the largest possible relative entropy
under upper bounds on the chi-squared divergences in both orders.
The sharp bound is attained by a pair of Bernoulli distributions
that meets both constraints with equality. Thus, the same bound
also solves the problem with exactly prescribed divergence values.

The two chi-squared divergences can have very different magnitudes.
For $P=(p,1-p)$ and $Q=(q,1-q)$ with $0<p,q<1$, they are
\[
 A=\frac{(p-q)^2}{q(1-q)},\qquad
 B=\frac{(p-q)^2}{p(1-p)}.
\]
Fix $a>0$ and set $p=q+\sqrt{aq(1-q)}$. For all sufficiently small
$q>0$, both distributions have full support, $A=a$, and
$B \sim \sqrt{aq}$ as $q \to 0^+$. For this family,
\[
 D(P\|Q) \sim -\tfrac12 \sqrt{aq} \log q \longrightarrow 0.
\]
The forward chi-squared divergence can also grow without bound
while the reverse chi-squared divergence tends to zero.
Taking $p=t$ and $q=t^3$ gives $A\sim \frac{1}{t}$ and $B\sim t$ as
$t \to 0^+$, while
\[
 D(P\|Q)\sim -2t\log t \longrightarrow 0.
\]

These examples also illustrate the limitations of local comparisons
between divergences. Csisz\'{a}r and Shields discuss the local
approximation of smooth $f$-divergences by chi-squared divergence
in~\cite[Theorem~4.1]{CsiszarS_FnT}, while
Sason~\cite{Sason2018} establishes more general results for probability
measures on general measurable spaces; see also \cite[Chapter~7.10]{PolyanskiyW25}.
In particular, for a convex function $f$ with $f(1)=0$ that is twice
differentiable at $1$ with $f''(1)>0$,
\begin{align}
\label{eq: local}
 D_f(P\|Q)\sim \tfrac12 f''(1) \, \chi^2(P\|Q)
\end{align}
as $P$ approaches a fixed full-support distribution $Q$ on
a finite alphabet. For relative entropy, this gives
$D(P\|Q)\sim \tfrac12(\log e)\chi^2(P\|Q)$.
Nielsen and Nock~\cite{NielsenNock2014} derive approximations 
by Taylor expanding the generator of an $f$-divergence.
In our notation, the quadratic term is the right-hand side of
\eqref{eq: local}, and higher-order terms involve moments of
$P(X)/Q(X)-1$ under $Q$.
As emphasized in~\cite[Remark~4.1]{CsiszarS_FnT},
this equivalence need not hold when $Q$ varies and its
probabilities approach zero. The examples above fall outside
the scope of this local approximation and motivate bounds
that account for the chi-squared divergences in both orders.

\paragraph{Related bounds.}
General comparisons among probability metrics and divergences
are surveyed by Gibbs and Su~\cite{GibbsSu02}, while Sason and
Verd\'u~\cite{SasonVerdu2016} develop a comprehensive treatment
of $f$-divergence inequalities.
A further motivation for comparing relative entropy and chi-squared
divergence is that both tensorize. In recent work, Cruz, Calmon,
and Yu~\cite[Theorem~2]{CruzCalmonYu2026} characterize the
$f$-divergences that tensorize. For the divergences considered here,
$D(P\|Q)$, $\log(1+\chi^2(P\|Q))$, and $\log(1+\chi^2(Q\|P))$
are each additive for product measures. Thus, the two ordered
chi-squared divergences of each factor suffice to evaluate our
bound for the product measures.

Comparisons between relative entropy and chi-squared divergence have
been studied through moment inequalities, integral representations,
and the general theory of $f$-divergences
\cite{DragomirG_01,NishiyamaS20,Sason2018,Simic2008,Simic2018}.
Of particular relevance here is the bound of Sason and
Verd\'u~\cite[Theorem~20]{SasonVerdu2016}. For $P\ne Q$ and finite
$A,B$, it gives
\begin{equation}
\label{eq1: 12.11.26}
 D(P\|Q)\leq \log(1+A)
 -\frac{3A^2\log e}{2\bigl((1+A)^2(1+B)-1\bigr)}.
\end{equation}
The right-hand side of \eqref{eq1: 12.11.26} extends continuously to 
zero at $A=B=0$. The correction term improves~\eqref{eq: well-known}, 
and the bound is locally tight under the convergence hypothesis of 
the cited theorem yielding \eqref{eq: local}. For fixed $A>0$, however, 
the right-hand side of \eqref{eq1: 12.11.26} tends to
$\log(1+A)-\frac{3A}{2(A+2)}\log e>0$ as $B \to 0^+$.
Thus this bound does not establish the vanishing of relative entropy
in that regime.

In recent independent work, Bari\'c, Iveli\'c Bradanovi\'c, and
Lovri\v{c}evi\'c~\cite[Example~1, Eq.~(42)]{BaricIvelicBradanovicLovricevic2026}
apply inequalities for the harmonic Jensen functional to obtain
two-sided bounds on relative entropy. Their upper bound involves
both ordered chi-squared divergences and a sum that depends on the
individual likelihood ratios. Here we ask how large $D(P\|Q)$ can
be when only $A$ and $B$ are known. For each $A,B>0$, we obtain
the smallest upper bound valid for every pair with those values
and exhibit a Bernoulli pair that attains it. The distinction also
appears in the lower bounds. Their lower bound depends on the smallest 
likelihood ratio and can be positive for a given pair. We show that, 
on any alphabet with at least three elements, $D(P\|Q)$ can be made 
arbitrarily small while both ordered chi-squared divergences remain 
fixed at their prescribed values $A$ and $B$.

Chi-squared comparisons and moment inequalities also arise in
bounds on information quantities. Tavakoli, Nguyen, and
Bose~\cite{TavakoliNguyenBose26} develop chi-squared bounds on
mutual information and channel dispersion, with applications
to finite-blocklength coding. Their analysis includes channels
whose transition probabilities are uniformly close, in relative
terms, to the output marginal.
Reeves~\cite{Reeves2020} derives bounds on R\'{e}nyi entropy from
two moments and bounds on mutual information through integrals
involving the variance of conditional densities.
In the setting of strong data-processing inequalities,
Raginsky~\cite{Raginsky2016} studies contraction coefficients
for $\Phi$-divergences, including relative entropy and
chi-squared divergence, and their connections to maximal
correlation and $\Phi$-Sobolev inequalities.
Polyanskiy and Wu~\cite{PolyanskiyW17} survey channel contraction
inequalities, including the equality of the contraction
coefficients for relative entropy and chi-squared divergence
when optimized over both input distributions.
Related applications of integral relations between relative
entropy and chi-squared divergence are developed by Nishiyama
and Sason~\cite{NishiyamaS20}, including bounds on contraction
coefficients and maximal correlation, as well as convergence
to stationarity of finite-state Markov chains.

The question of controlling relative entropy without uniformly
bounded likelihood ratios also arises in statistical applications.
Kaji~\cite{Kaji2026} characterizes necessary and sufficient
conditions for Hellinger bounds on the Kullback--Leibler
divergence and related log-likelihood quantities, allowing
unbounded likelihood ratios. In the present problem, the two
ordered chi-squared constraints bound moments of the
likelihood ratio, and we determine the resulting sharp upper
bound on relative entropy without imposing a uniform bound
on that ratio.

Related comparisons have been developed through skew divergences
and convexity methods. Melbourne, Madiman, and
Salapaka~\cite{MelbourneMadimanSalapaka2019} establish
differential relationships between skew relative entropy and
a skew chi-squared divergence that interpolates between
$\chi^2(P\|Q)$ and $\chi^2(Q\|P)$. Melbourne~\cite{Melbourne2020}
studies how strong convexity of divergence generators yields
comparisons with chi-squared divergence and refinements of
classical divergence inequalities. Melbourne
et al.~\cite{MelbourneEtAl2022} use skew-divergence methods
to obtain bounds on the entropy concavity deficit of mixtures,
with applications to mutual information.

The problem also belongs to the theory of sharp comparisons among
$f$-divergences. Harremo\"es and Vajda~\cite{HarremoesVajda2011}
study the joint range of two $f$-divergences; see also 
\cite[Chapter~7.4]{PolyanskiyW25}. On a finite alphabet, 
Prelov~\cite{Prelov2019} determined the maximum of each of
$D(P\|Q)$ and $D(Q\|P)$ for fixed $Q$ and prescribed variational
distance between $P$ and $Q$. For $D(P\|Q)$, he also obtained the
exact maximum when only the smallest mass of $Q$ and the variational
distance are prescribed. Prelov~\cite{Prelov2020} subsequently
extended the analysis with one distribution fixed to $f$-divergences
and R\'{e}nyi divergence.  A related bound with a fixed uniform 
reference distribution follows from Cicalese, Gargano, and
Vaccaro~\cite{CicaleseGarganoVaccaro2018}. For a distribution $P$
with full support on $m$ points, they lower bound $H(P)$ in terms
of $m$ and an upper bound on the ratio of its largest to its
smallest mass. Since $D(P\|U_m)=\log m-H(P)$, where $U_m$ is
uniform on these points, their result gives an upper bound on
relative entropy when the ratio of the largest to the smallest
value of $P/U_m$ is bounded. In the present problem, both 
distributions vary, and we maximize relative entropy subject 
to upper bounds on the two ordered chi-squared divergences.

For an $f$-divergence subject to upper bounds on two other
$f$-divergences, Guntuboyina, Saha, and
Schiebinger~\cite[Theorem~2.1]{GuntuboyinaSahaSchiebinger2014}
show that restricting the alphabet to at most four points leaves
the supremum unchanged. This reduction also applies when the
objective is total variation, which determines the minimum error
probability in binary hypothesis testing with equal priors.
For the relative entropy problem considered here, we show that a
Bernoulli pair attains the supremum and meets both chi-squared
constraints with equality. Binary extremizers also arise in sharp
lower bounds on divergences under prescribed means and
variances~\cite{NishiyamaS20,Nishiyama2021}.

The two constraint generators are $f_1(t)=(t-1)^2$ and
$f_2(t)=(t-1)^2/t$ for $t>0$, corresponding to the forward and
reverse chi-squared divergences, respectively. These generators
give a formulation as a generalized moment problem, a classical
framework treated in~\cite{KreinNudelman1977}. Indeed, under $Q$,
the likelihood ratio $L=P/Q$ satisfies
\[
 \mathbb E_Q[L]=1,\qquad
 \mathbb E_Q[L^2]=1+A,\qquad
 \mathbb E_Q[L^{-1}]=1+B,
\]
and the objective is $\mathbb E_Q[L\log L]$. We show that a likelihood 
ratio taking only two values attains the upper bound.

\paragraph{Main results.}
For $a,b>0$, let
\[
 F(a,b):=\sup_{P,Q}\bigl\{D(P\|Q):
           \chi^2(P\|Q)\leq a,\ \chi^2(Q\|P)\leq b\bigr\}.
\]
We give the following explicit expression for $F$: 
\begin{align}
\label{eq: function F}
F(a,b) = \frac12 \log \frac{a}{b}
+\frac{ab+b-a}{\sqrt{(a+b+ab)^2-4ab}} \; 
\log\left( \frac{a+b+ab+\sqrt{(a+b+ab)^2-4ab}}{2\sqrt{ab}} \right), \quad a,b>0,
\end{align}
and construct a pair of
Bernoulli distributions that simultaneously attains
\[
 D(P\|Q)=F(a,b),\qquad D(Q\|P)=F(b,a),
\]
with $\chi^2(P\|Q)=a$ and $\chi^2(Q\|P)=b$. Interchanging the
two alphabet symbols gives an equivalent Bernoulli pair with the 
labels interchanged. Thus prescribing the two divergence values 
exactly gives the same optimum as imposing them as upper bounds.
We set $F(0,0)=0$, consistent with the optimization problem and 
the continuous extension of~\eqref{eq: function F} at the origin.

As a consequence, if one of the two ordered chi-squared divergences
tends to zero along a sequence of pairs $(P_n,Q_n)$ and the other
remains uniformly bounded, then both $D(P_n\|Q_n)$ and
$D(Q_n\|P_n)$ tend to zero. The boundedness hypothesis cannot be 
dropped if convergence of both relative entropies is required.
For example, take Bernoulli parameters $p=t$ and $q=t c^{-1/t}$. 
As $t \to 0^+$, $\chi^2(Q\|P)\sim t$, whereas $D(P\|Q)\to1$ 
in the chosen logarithmic base. Indeed, $p\log(p/q)=1$ and the 
contribution of the other symbol tends to zero. In this example, 
$\chi^2(P\|Q)\to\infty$, while $D(Q\|P)\to0$ by~\eqref{eq: well-known} 
with $P$ and $Q$ interchanged.

An application concerns mutual information and lautum information.
The chi-squared mutual information between random variables $X$ and $Y$
is defined as
\[
I_{\chi^2}(X;Y)
:= \chi^2\left(P_{XY}\,\middle\|\,P_XP_Y\right),
\]
which provides a chi-squared analogue of the mutual information
$I(X;Y)=D(P_{XY}\|P_X P_Y)$; it has been studied, for example,
in connection with subadditivity properties for graphical
models~\cite{AbbeBoixAdsera2019}.
It also coincides with twice the squared-loss mutual information under
the normalization adopted in~\cite{Sugiyama2013}.
In the reverse direction, we define the \emph{chi-squared lautum information}
as
\[
L_{\chi^2}(X;Y)
:= \chi^2\left(P_X P_Y\,\middle\|\,P_{XY}\right),
\]
by analogy with the lautum information
$L(X;Y)=D(P_XP_Y\|P_{XY})$ introduced by Palomar and
Verd\'u~\cite{PalomarVerdu2008}.
Here, ``chi-squared lautum information'' is terminology adopted
in this work, rather than a widely established name.
Both chi-squared quantities are nonnegative and vanish if and only if
$X$ and $Y$ are independent, but they generally differ because
the chi-squared divergence is asymmetric. When both quantities 
are finite and positive, our main result gives
\begin{align}
\label{eq: MI}
I(X;Y) &\leq F\bigl(I_{\chi^2}(X;Y),L_{\chi^2}(X;Y)\bigr), \\
\label{eq: LI}
L(X;Y) &\leq F\bigl(L_{\chi^2}(X;Y),I_{\chi^2}(X;Y)\bigr).
\end{align}

In contrast to this attained maximum, on any finite or countably
infinite alphabet containing at least three elements, the infimum
of $D(P\|Q)$ under the exact constraints
$\chi^2(P\|Q)=a$ and $\chi^2(Q\|P)=b$ is zero for every $a,b>0$.
This infimum is not attained, since $D(P\|Q)=0$ implies $P=Q$,
which would force both chi-squared divergences to vanish.
On a binary alphabet, however, these exact constraints determine
$D(P\|Q)$ uniquely as $F(a,b)$, so the minimum and maximum coincide.
Thus, allowing a third alphabet symbol lowers the infimum
from the strictly positive value $F(a,b)$ to zero.

We apply the sharp upper bound on the relative entropy to the analysis 
of redundancy reduction by mixing in lossless source coding.
Let $P^n$ and $Q^n$ denote the product measures
induced by $P$ and $Q$, respectively, on blocks of length $n$. When
$P^n$ generates the block, replacing $Q^n$ with the mixture
$R_\lambda^{(n)}:=\lambda P^n+(1-\lambda)Q^n$ yields an expected
saving in ideal codelength of
\[
  \Delta_\lambda(P^n,Q^n)
  = D(P^n\|Q^n)-D(P^n\|R_\lambda^{(n)}).
\]
Proposition~\ref{proposition: bounds on Delta} gives a lower bound
on this saving in terms of $D(P\|Q)$ and an upper bound in terms of
$F$ and the two ordered chi-squared divergences between $P$ and $Q$.
Thus, both bounds can be evaluated from single-letter quantities.
For each fixed $n$, the upper bound agrees with the saving to first
order as $\lambda\to0^+$: both have the same right derivative at zero,
namely $\bigl[(1+\chi^2(P\|Q))^n-1\bigr]\log e$.

The paper is organized as follows. 
Section~\ref{section: Infimum of KL with Prescribed Forward and Reverse Chi-Square Divergences} 
briefly examines the infimum of the relative entropy under prescribed forward and reverse 
chi-squared divergences. Section~\ref{section: attainable upper bound on the relative entropy}
presents the sharp upper bound, its Bernoulli extremizers, and its asymptotic properties, 
including small-parameter scalings. Finally, 
Section~\ref{section: redundancy reduction by mixing in lossless source coding}
derives the bounds on redundancy reduction by mixing and establishes the first-order 
tightness of the upper bound.

\section{The infimum of relative entropy with prescribed forward and reverse chi-squared divergences}
\label{section: Infimum of KL with Prescribed Forward and Reverse Chi-Square Divergences}

The following proposition shows that, for any prescribed positive values
of the forward and reverse chi-squared divergences, the infimum of
relative entropy is zero on any alphabet with at least three points.
On a two-point alphabet, the two prescribed values uniquely determine
the relative entropy, which is strictly positive.

The general joint-range theorem of Guntuboyina, Saha, and Schiebinger
\cite[Theorem~4.1]{GuntuboyinaSahaSchiebinger2014} shows that, for any
collection of $m$~$f$-divergences with $m \geq 1$, every point in their
joint range can be realized by a pair of distributions on an alphabet
of at most $m+2$ points. For the three divergences considered here,
this gives a bound of five points for realizing any attainable triple.
We are unaware of an explicit statement in the literature that three
points suffice to make the relative entropy arbitrarily close to zero
while both chi-squared divergences remain fixed at their prescribed
positive values. We give a self-contained three-point construction.

\begin{proposition}
\label{proposition: infimum of KL}
Let $\mathcal X$ be a finite or countably infinite set containing at
least three elements, and let $A,B>0$. Then
\[
\inf\bigl\{D(P\|Q):\ \chi^2(P\|Q)=A,\quad
                      \chi^2(Q\|P)=B\bigr\}=0,
\]
where the infimum ranges over pairs of probability mass functions
$P$ and $Q$ on $\mathcal X$ satisfying the displayed constraints.
The infimum is not attained, but it can be approached by pairs 
supported on any fixed three-element subset of $\mathcal X$.
\end{proposition}

\begin{proof}
Since $A>0$, every admissible pair satisfies $P\ne Q$ and hence
$D(P\|Q)>0$. To show that the infimum is zero, fix any three elements
of $\mathcal X$ and label them $0,1,2$. It suffices to construct the
desired pairs on $\{0,1,2\}$, since extending both mass functions by
zero elsewhere preserves all three divergences.

For each sufficiently large $r$, we have $r>\max\bigl\{1,\frac{A}{B}\bigr\}$ 
and
\[
(r-1)\left(1-\frac{A}{Br}\right)>A,
\]
since the left-hand side tends to infinity as $r\to\infty$. Set
\[
s=\frac{A}{Br},\qquad \delta=\frac{A}{r-s},\qquad
c_r=1-\frac{\delta}{r-1}-\frac{\delta}{1-s}.
\]
Then $0<s<1<r$ and $\delta>0$. Moreover,
\[
c_r
=1-\frac{\delta(r-s)}{(r-1)(1-s)}
=1-\frac{A}{(r-1)(1-s)}>0.
\]
Define
\[
\begin{array}{c|ccc}
x&0&1&2\\ \hline \\[-10pt]
Q_r(x)&c_r&\dfrac{\delta}{r-1}&\dfrac{\delta}{1-s}\\[10pt]
P_r(x)&c_r&\dfrac{r\delta}{r-1}&\dfrac{s\delta}{1-s}
\end{array}
\]
and assign zero mass to any remaining elements of $\mathcal X$.
The entries in each row are nonnegative and sum to one; indeed,
$P_r(1)-Q_r(1)=\delta$ and $P_r(2)-Q_r(2)=-\delta$.
Consequently,
\begin{align*}
\chi^2(P_r\|Q_r)
&=\frac{\delta^2}{Q_r(1)}+\frac{\delta^2}{Q_r(2)} \\
&=\delta(r-1)+\delta(1-s) \\
&=A,
\end{align*}
and 
\begin{align*}
\chi^2(Q_r\|P_r)
&=\frac{\delta^2}{P_r(1)}+\frac{\delta^2}{P_r(2)} \\
&=\delta \; \frac{r-1}{r}+\delta \; \frac{1-s}{s} \\
&=\frac{\delta(r-s)}{rs} \\
&=B.
\end{align*}
On the other hand, the likelihood ratios at $0,1,2$ are $1,r,s$,
respectively, and hence
\[
D(P_r\|Q_r)
=\frac{r\delta}{r-1} \; \log r
 +\frac{s\delta}{1-s} \; \log s.
\]
Since $\log s<0$, nonnegativity of relative entropy yields
\[
0\le D(P_r\|Q_r)
\le \frac{Ar\log r}{(r-s)(r-1)}
\underset{r \to \infty}{\longrightarrow} 0.
\]
Thus the admissible pairs $(P_r,Q_r)$, all supported on the chosen
three-element subset of $\mathcal X$, approach the asserted infimum.
\end{proof}

\begin{remark}[The distinction between binary alphabets and larger alphabets]
{\em
The assumption that $\mathcal X$ contains at least three elements is
essential. On a two-element alphabet, the constraints $\chi^2(P\|Q)=A$
and $\chi^2(Q\|P)=B$, with $A,B>0$, determine the pair $(P,Q)$ uniquely
up to a simultaneous relabeling of the two elements. They therefore
determine a strictly positive value of $D(P\|Q)$.

To see this, note that the finiteness of both chi-squared divergences implies 
that $P$ and $Q$ have the same support. Since $A>0$, they are distinct,
so their common support contains both elements. The likelihood ratio
$P/Q$ has mean one under $Q$ and is nonconstant. After relabeling, its
two values can therefore be written as $r>1$ and $0<s<1$.
Normalization gives
\begin{align*}
P&=\left(\frac{r(1-s)}{r-s},\,\frac{s(r-1)}{r-s}\right),\\[0.15cm]
Q&=\left(\frac{1-s}{r-s},\,\frac{r-1}{r-s}\right).
\end{align*}
Direct calculation yields
\[
A=(r-1)(1-s),\qquad B=\frac{(r-1)(1-s)}{rs}.
\]
Setting $c=\frac{A}{B}$, we therefore obtain
\[
rs=c,\qquad r+s=1+A+c.
\]
Thus $r$ and $s$ are the roots of $t^2-(1+A+c)t+c$. This polynomial
is positive at $0$, negative at $1$, and tends to infinity as
$t\to\infty$. Hence its roots satisfy $0<s<1<r$ and are given by
\begin{align*}
r&=\frac{1+A+c+\sqrt{(1+A+c)^2-4c}}{2},\\[0.15cm]
s&=\frac{1+A+c-\sqrt{(1+A+c)^2-4c}}{2}.
\end{align*}
Conversely, for every $A,B>0$, substituting these roots into the
formulas for $P$ and $Q$ gives a pair with the prescribed divergences.
The unique value of the relative entropy is therefore
\[
D(P\|Q)
=\frac{r(1-s)\log r+s(r-1)\log s}{r-s}>0.
\]
In particular, the binary infimum is strictly positive and is attained.

In the construction used in the proof of Proposition~\ref{proposition: infimum of KL}, 
the binary identity $A=(r-1)(1-s)$ is equivalent to $c_r=0$.
On three points, the common mass $c_r$ at the additional point allows 
$r$ to grow arbitrarily large while both chi-square divergences retain 
their prescribed values.

A similar distinction between binary alphabets and alphabets with at
least three elements arises in Fischer's characterization of logarithmic
functions by a Shannon-type inequality~\cite{Fischer1972}
(see also \cite[Section~4.3]{AczelD1975}). 
Jiao {\em et al.}~\cite[Sections~III-A--III-C]{Jiao2014} discuss
Fischer's result and examine three other questions about information
divergences, showing in each case that the answer depends on whether
the alphabet has two elements or at least three.
}
\end{remark}

\section{An attainable upper bound on the relative entropy with 
prescribed forward and reverse chi-squared divergences}
\label{section: attainable upper bound on the relative entropy}

How large can the relative entropy \(D(P\|Q)\) be when the forward and
reverse chi-squared divergences, \(\chi^2(P\|Q)\) and \(\chi^2(Q\|P)\),
are prescribed? In Section~\ref{subsection: UB on KL}, we derive an upper
bound on \(D(P\|Q)\) in terms of these two values. In
Section~\ref{subsection: KL between binomial distributions}, we determine
the maximum relative entropy between binomial distributions under the
same constraints and examine its properties. In
Section~\ref{subsection: attainability}, we show that, for any prescribed
pair of divergence values, two pairs of probability mass functions on a binary
alphabet attain the upper bound. Finally, in Section~\ref{subsection: asymptotic}, 
we study the asymptotic properties and small-parameter scalings of the attainable 
upper bound. 

\subsection{Upper bound on the relative entropy expressed in terms of chi-squared divergences}
\label{subsection: UB on KL}

The following result gives the upper bound on the relative entropy. 

\begin{proposition}
\label{proposition: KL and chi^2}
Let \(P\) and \(Q\) be distinct probability mass functions on a finite or
countably infinite alphabet \(\mathcal X\) such that
\(A := \chi^2(P\|Q) < \infty\) and \(B := \chi^2(Q\|P) < \infty\).
Then 
\begin{align}
\label{eq1: 11.11.26}
D(P\|Q) &\leq F(A,B) \\
\label{eq2: 11.11.26}
& < (\tfrac13 A+\tfrac16 B) \, \log e,
\end{align}
where the function $F \colon (0, \infty)^2 \to (0, \infty)$ is given in 
\eqref{eq: function F}. These bounds have the following properties:
\begin{enumerate}
\item 
The upper bound on the right-hand side of \eqref{eq1: 11.11.26}
uniformly improves upon the bound 
\begin{align}
\label{eq6: 11.11.26}
D(P\|Q) \leq \log(1+A).
\end{align}
For each fixed \(A>0\), the supremum of the right-hand side of
\eqref{eq1: 11.11.26} over \(B>0\) equals \(\log(1+A)\).

\item 
The upper bound on the right-hand side of \eqref{eq1: 11.11.26}
also uniformly improves upon the upper bound on the right-hand side
of \eqref{eq1: 12.11.26}.

\item 
Let \(G(A,B)\) and \(F(A,B)\) denote the upper bounds on \(D(P\|Q)\)
given by the right-hand sides of \eqref{eq1: 12.11.26} and
\eqref{eq1: 11.11.26}, respectively. For each fixed \(A>0\), define
\begin{align}
\label{eq: g function}
g(A)=\log(1+A)-\frac{3A}{2(A+2)} \; \log e.
\end{align}
Then \(g(A)>0\), and the following expansions hold as
\(\varepsilon \to 0^+\):
\begin{align}
F(A,\varepsilon)
 &=\varepsilon \log\left(\frac A{\varepsilon e} \right)
 +O\left(\varepsilon^2\log\frac1\varepsilon\right),
 \label{eq10: 12.11.26}\\[0.1cm]
G(A,\varepsilon)
 &=g(A)+\frac{3(A+1)^2 \log e}{2(A+2)^2} \cdot \varepsilon
 +O(\varepsilon^2).
 \label{eq11: 12.11.26}
\end{align}
Consequently,
\begin{gather}
\lim_{B \to 0^+} F(A,B) = 0,\qquad
\lim_{B \to 0^+} G(A,B) = g(A) > 0,
\label{eq12: 12.11.26}\\
\frac{G(A,\varepsilon)}{F(A,\varepsilon)}
\sim \frac{g(A)}{\varepsilon \log(1/\varepsilon)}
\longrightarrow +\infty,
\label{eq13: 12.11.26}
\end{gather}
and
\begin{equation}
\lim_{A \to \infty} \lim_{B \to 0^+} G(A,B)=+\infty,
\qquad
\lim_{A \to \infty} \lim_{B \to 0^+} F(A,B)=0.
\label{eq14: 12.11.26}
\end{equation}

\item 
Let \(\{P_n\}\) be a sequence of probability mass functions and let
\(Q\) be a probability mass function, all on the same finite or
countably infinite alphabet \(\mathcal X\). Assume that \(P_n \ne Q\) 
and \(P_n \ll Q\) for all sufficiently large \(n\) and that 
\( \frac{P_n}{Q} \) converges  uniformly to \(1\) on the support of \(Q\), i.e.,
\begin{align}
\label{eq: uniform convergence of LR}
\lim_{n \to \infty} \sup_{x: \, Q(x)>0} \left| \frac{P_n(x)}{Q(x)} - 1 \right| = 0.
\end{align}
Set \(A_n := \chi^2(P_n\|Q)\) and \(B_n := \chi^2(Q\|P_n)\), and suppose
that both are finite for all sufficiently large \(n\). Then the upper
bounds on \(D(P_n\|Q)\) on the right-hand sides of
\eqref{eq1: 11.11.26} and \eqref{eq2: 11.11.26} are locally tight
in the sense that
\begin{equation}
\label{eq2: 13.11.26}
\lim_{n \to \infty}
\frac{D(P_n \| Q)}{\bigl(\tfrac13 A_n+\tfrac16 B_n \bigr) \, \log e}
= 1.
\end{equation}
\end{enumerate}
\end{proposition}
The proof of Proposition~\ref{proposition: KL and chi^2} is given
later in this subsection.

\begin{remark}
{\em In Section~\ref{subsection: attainability}, we show that for every \(A,B>0\),
there exist two pairs of probability mass functions \((P,Q)\) on a binary
alphabet, each of which satisfies \(\chi^2(P\|Q)=A\) and \(\chi^2(Q\|P)=B\)
and attains the upper bound in \eqref{eq1: 11.11.26}.}
\end{remark}

\begin{corollary}
\label{corollary1: KL and chi^2}
Let \(\{P_n\}\) and \(\{Q_n\}\) be sequences of probability mass
functions on the same finite or countably infinite alphabet
\(\mathcal X\). Suppose that, for some constant \(a>0\) and some
integer \(n_0\),
\begin{align}
\chi^2(P_n\|Q_n) \leq a \quad \text{for all } n\geq n_0,
\qquad
\lim_{n \to \infty} \chi^2(Q_n\|P_n)=0.
\end{align}
Then
\begin{align}
\lim_{n \to \infty} D(P_n\|Q_n)=0,
\qquad
\lim_{n \to \infty} D(Q_n\|P_n)=0.
\end{align}
In other words, if one of the two chi-squared divergences is
eventually bounded by a finite constant, however large, and the
other converges to zero, then the relative entropies in both
directions converge to zero.
\end{corollary}

\begin{proof}
Set \(\varepsilon_n:=\chi^2(Q_n\|P_n)\). By assumption,
\(\varepsilon_n\to 0\). At every index for which \(\varepsilon_n=0\),
we have \(P_n=Q_n\), so both relative entropies vanish.

By Lemma~\ref{lem:integral}, which is presented later in this 
subsection for proving Proposition~\ref{proposition: KL and chi^2}, 
the expression for \(F(A,B)\) on the right-hand side of 
\eqref{eq1: 11.11.26} can be written as
\[
F(A,B)
= (\log e)\int_0^1
\frac{\lambda}
{(1-\lambda)/B+\lambda/A+\lambda(1-\lambda)}
\,\mathrm{d}\lambda.
\]
For every \(\lambda\in(0,1)\), the denominator decreases strictly
as either \(A\) or \(B\) increases with the other held fixed.
Hence \(F(A,B)\) is strictly increasing in each argument.

For all \(n \geq n_0 \) with \(\varepsilon_n>0\), the
bound on the right-hand side of \eqref{eq1: 11.11.26} and the
monotonicity of \(F\) give
\[
0\leq D(P_n\|Q_n)
\leq F\bigl(\chi^2(P_n\|Q_n),\varepsilon_n\bigr)
\leq F(a,\varepsilon_n).
\]
Since \(F(a,\varepsilon)\to 0\) as \(\varepsilon \to 0^+\)
by \eqref{eq10: 12.11.26}, it follows that \(D(P_n\|Q_n)\to 0\).

Finally, applying \eqref{eq: well-known} with \(P_n\) and \(Q_n\)
interchanged gives 
\[
0\leq D(Q_n\|P_n)\leq \log(1+\varepsilon_n).
\]
Hence \(D(Q_n\|P_n) \to 0\) as well.
\end{proof}

\begin{remark}
\label{remark: on corollary 1}
{\em Corollary~\ref{corollary1: KL and chi^2} cannot be deduced
from the upper bound on the right-hand side of \eqref{eq1: 12.11.26}.
Denote this upper bound on \(D(P\|Q)\) by \(G(A,B)\), where
\(A:=\chi^2(P\|Q)\) and \(B:=\chi^2(Q\|P)\). For finite \(A,B>0\),
\[
0<G(A,B)-g(A)
=\tfrac32 B\log e\,
 \left(\frac{1+A}{2+A}\right)^2
 \frac{A(A+2)}{A(A+2)+(1+A)^2B}
<\tfrac32 B\log e,
\]
where \(g\) is defined in \eqref{eq: g function}.
Since \(g(0)=0\) and
\(g'(x)=\frac{(x^2+x+1)\log e}{(1+x)(x+2)^2}>0\) for every \(x>0\),
it follows that \( g \) is strictly increasing on \( [0, \infty) \).
Thus, for \(a>0\),
\[\underset{0<A\leq a}{\sup} g(A)=g(a)>g(0)=0,\] and the preceding estimate gives
\[
0<g(a)\leq \sup_{0<A\leq a}G(A,B)
\leq g(a)+\tfrac32 B\log e.
\]
Thus, as \(B\to 0^+\), the supremum tends to \(g(a)\).

Under the hypotheses of Corollary~\ref{corollary1: KL and chi^2},
\(A_n:=\chi^2(P_n\|Q_n)\leq a\) for all sufficiently large \(n\), and
\(B_n:=\chi^2(Q_n\|P_n)\xrightarrow[n\to\infty]{}0\).
The best uniform asymptotic estimate furnished by
\eqref{eq1: 12.11.26} is therefore
\[
\limsup_{n\to\infty} D(P_n\|Q_n)\leq g(a).
\]
Since \(g(a)>0\), this estimate does not imply that \(D(P_n\|Q_n)\) 
tends to zero. Corollary~\ref{corollary1: KL and chi^2}, in contrast,
asserts this convergence under the same hypotheses.}
\end{remark}

\begin{remark}
\label{remark: counterexample}
{\em The eventual boundedness of \(\chi^2(P_n\|Q_n)\) in
Corollary~\ref{corollary1: KL and chi^2} cannot be omitted
if one wishes to conclude that \(D(P_n\|Q_n)\to 0\). Indeed, 
let $(t_n)_{n \geq 1}$ be a sequence in $(0,1)$ such that $t_n \to 0$, 
and define the two sequences of probability mass functions 
\( \{ P_n \}_{n \geq 1} \) and \( \{ Q_n \}_{n \geq 1} \) on the binary 
alphabet by 
\[
P_n=(t_n,1-t_n),
\qquad
Q_n=\bigl(t_n e^{-1/t_n},\,1-t_n e^{-1/t_n}\bigr).
\]
Then, as $n\to\infty$,
\[
\chi^2(Q_n\|P_n)
=\frac{t_n\bigl(1-e^{-1/t_n}\bigr)^2}{1-t_n}
\sim t_n \longrightarrow 0,
\]
and
\[
\chi^2(P_n\|Q_n)
=\frac{t_n e^{1/t_n}\bigl(1-e^{-1/t_n}\bigr)^2}
       {1-t_n e^{-1/t_n}}
\sim t_n e^{1/t_n}\longrightarrow\infty.
\]
With natural logarithms, the first coordinate contributes exactly
\(t_n\log\bigl(t_n/(t_n e^{-1/t_n})\bigr)=1\) to \(D(P_n\|Q_n)\).
Hence
\[
D(P_n\|Q_n)
=1+(1-t_n)\log\left(
  \frac{1-t_n}{1-t_n e^{-1/t_n}}
\right)
=1-t_n+\frac{t_n^2}{2}+O(t_n^3)
\longrightarrow 1.
\]
Thus, even on a binary alphabet, \(\chi^2(Q_n\|P_n)\to 0\) alone
does not imply that \(D(P_n\|Q_n)\to 0\). Nevertheless, interchanging 
\(P\) and \(Q\) in \eqref{eq: well-known} gives
\[
0\leq D(Q_n\|P_n)
\leq \log\bigl(1+\chi^2(Q_n\|P_n)\bigr)
\xrightarrow[n\to\infty]{}0.
\]}
\end{remark}

\begin{example}
\label{example: numerical}
{\em Let \(P\) and \(Q\) be probability mass functions on a finite or
countably infinite alphabet, and suppose that 
\[ \chi^2(P \| Q) \leq 100, \qquad \chi^2(Q \| P) \leq \tfrac1{100}. \]

\begin{table}[hbt]
    \centering
    \caption{Comparison of upper bounds on the ordered relative entropies in nats (natural logarithms).}
    \label{tab:upper bounds}
    \vspace{6pt}
    \begin{tabular}{|l|c|c|}
        \hline
              & \multicolumn{2}{c|}{Upper bounds on} \\ 
              & \( D(P \| Q) \) & \( D(Q \| P) \) \\ \hline 
        Eq.~\eqref{eq: well-known} & $ 4.615120521 $       & $ 0.00995033 $      \\ \hline      
        Eq.~\eqref{eq1: 12.11.26}  & $ 3.159093975 $       & $ 0.00994886 $      \\ \hline
        Eq.~\eqref{eq1: 11.11.26}  & $ \bf{0.08145515} $   & $ \bf{0.00994226} $ \\ \hline
        Eq.~\eqref{eq2: 11.11.26}  & $33.335000000$        & $ 16.67000000 $     \\ \hline
    \end{tabular}
\end{table}

Table~\ref{tab:upper bounds} compares the upper bounds on the ordered relative
entropies \( D(P \| Q) \) and \( D(Q \| P) \).
The boldface row, corresponding to the bounds in Eq.~\eqref{eq1: 11.11.26},
shows that both ordered relative entropies are small. This is consistent 
with Corollary~\ref{corollary1: KL and chi^2}. In contrast, the bounds
in Eqs.~\eqref{eq1: 12.11.26}, \eqref{eq: well-known},
and~\eqref{eq2: 11.11.26} establish that \( D(Q \| P) \) is small,
but do not establish the same for \( D(P \| Q) \).
Furthermore, Proposition~\ref{proposition: KL and chi^2} shows that
the bounds in Eq.~\eqref{eq1: 11.11.26} are the tightest among those
compared for both ordered relative entropies.
}
\end{example}

\begin{corollary}
\label{corollary2: KL and chi^2}
Let \(P\) and \(Q\) be distinct probability mass functions on a finite or
countably infinite alphabet \(\mathcal X\), and suppose that 
\( A := \chi^2(P \| Q) < \infty \) and \( B := \chi^2(Q \| P) < \infty \). Then, 
\begin{align}
& \hspace*{-0.3cm} D(P\|Q) + D(Q\|P) \nonumber \\
\label{eq2: 12.11.26}
& \hspace*{-0.1cm} \leq \frac{2AB}{\sqrt{(A+B+AB)^2-4AB}} \; 
\log\left( \frac{A+B+AB+\sqrt{(A+B+AB)^2-4AB}}{2\sqrt{AB}} \right) \\
\label{eq3: 12.11.26}
& \hspace*{-0.1cm} < \tfrac12 (A+B) \, \log e.
\end{align}
\end{corollary}
\begin{proof}
Applying \eqref{eq1: 11.11.26} separately to \(D(P \| Q)\) and \(D(Q \| P)\) gives
\[
D(P \| Q) + D(Q \| P) \leq F(A,B) + F(B,A).
\]
By the definition of \(F\) in \eqref{eq: function F}, the right-hand side equals
that of \eqref{eq2: 12.11.26}, proving \eqref{eq2: 12.11.26}. Applying
\eqref{eq2: 11.11.26} to each term gives 
\begin{align*}
F(A,B) + F(B,A)
& \leq \left(\tfrac13 A + \tfrac16 B\right) \log e 
   + \left(\tfrac13 B + \tfrac16 A\right) \log e  \\
&= \tfrac12(A+B) \log e,
\end{align*}
which proves \eqref{eq3: 12.11.26}.
\end{proof}

\begin{remark}
{\em By \eqref{eq3: 12.11.26}, the upper bound on the right-hand side of \eqref{eq2: 12.11.26}
sharpens the inequality \cite{Simic2008}
\[
D(P\|Q)+D(Q\|P) \leq \tfrac{1}{2}
\bigl(\chi^2(P\|Q)+\chi^2(Q\|P)\bigr)\log e.
\]
In Section~\ref{subsection: attainability}, we show that for every \(A,B>0\),
there exist two pairs of probability mass functions \((P,Q)\) on a binary
alphabet, each of which satisfies \(\chi^2(P\|Q)=A\) and \(\chi^2(Q\|P)=B\)
and attains the upper bound in \eqref{eq2: 12.11.26}.}
\end{remark}

We next prove Proposition~\ref{proposition: KL and chi^2} by first introducing and proving four lemmata. 
\begin{lemma}
\label{lemma: UB by chi squares}
Let \(P\) and \(Q\) be probability mass functions on a finite or
countable alphabet \(\mathcal X\), with a common support. 
Set \(A=\chi^2(P\|Q)\) and \(B=\chi^2(Q\|P)\), and let
\[
R_\lambda(x):=(1-\lambda)P(x)+\lambda Q(x), \qquad \lambda\in[0,1].
\]
If \(0<A,B<\infty\), then
\begin{equation}\label{eq:bound}
\sum_{x\in\mathcal X}\frac{(P(x)-Q(x))^2}{R_\lambda(x)}
\leq
\frac{AB}{(1-\lambda)A+\lambda B+\lambda(1-\lambda)AB}.
\end{equation}
The bound extends to all \(P\) and \(Q\) with common support as follows.
If \(P=Q\), both sides of \eqref{eq:bound} are taken to be zero.
If \(P\ne Q\) and at least one of \(A\) and \(B\) is infinite, the
right-hand side of \eqref{eq:bound} for \(0<\lambda<1\) is given by
\begin{equation}\label{eq:extended}
\left(\frac{1-\lambda}{B}+\frac{\lambda}{A}
      +\lambda(1-\lambda)\right)^{-1},
\end{equation}
with the convention \( \frac{1}{\infty}=0 \).
At \(\lambda=0\) and \(\lambda=1\), equality holds with common values
\(B\) and \(A\), respectively, including infinite values.
\end{lemma}

\begin{proof}
All sums below are taken over the common support. Points where
\(P(x)=Q(x)=0\) contribute zero and are omitted.
If \(P=Q\), the assertion is immediate. At \(\lambda=0\), the
denominator on the left-hand side of \eqref{eq:bound} is \(P(x)\),
so the sum equals \(B\). Similarly, at \(\lambda=1\), the sum
equals \(A\). It therefore suffices to consider \(P\ne Q\) and
\(0<\lambda<1\).

For brevity, write \(\Delta(x):=P(x)-Q(x)\), and let
\begin{align}
\label{eq0: 23.11.26}
T_\lambda:=\sum_{x\in\mathcal X}\frac{\Delta(x)^2}{R_\lambda(x)},
\qquad
t:=\lambda(1-\lambda).
\end{align}
Since \(|\Delta(x)|\leq P(x)+Q(x)\) and
\(R_\lambda(x)\geq\min\{\lambda,1-\lambda\}(P(x)+Q(x))\), we have
\[
\frac{\Delta(x)^2}{R_\lambda(x)}
\leq\frac{P(x)+Q(x)}{\min\{\lambda,1-\lambda\}}.
\]
Summing both sides over \(x\in\mathcal X\) gives
\(T_\lambda \leq \frac{2}{\min\{\lambda,1-\lambda\}} < \infty\).

Define the weighted inner product
\begin{align}
\label{eq1: 23.11.26}
\langle f,g \rangle_\lambda
:=\sum_{x\in\mathcal X} \frac{P(x) \, Q(x)}{R_\lambda(x)} \; f(x)g(x),
\end{align}
and consider the functions
\begin{align}
\label{eq2: 23.11.26}
f_0=1,\qquad
f_1=\lambda \, \frac{\Delta}{P},\qquad
f_2=-(1-\lambda) \, \frac{\Delta}{Q}.
\end{align}
We use \(f_0\) in all cases, \(f_1\) whenever \(B<\infty\), and
\(f_2\) whenever \(A<\infty\). To see that each retained function
belongs to the corresponding weighted \(\ell^2\) space, use
\(R_\lambda\geq(1-\lambda)P\) and \(R_\lambda\geq\lambda Q\) to obtain
\begin{subequations}\label{eq:weighted-norm-bounds}
\begin{align}
\langle f_0,f_0 \rangle_\lambda
&=\sum_{x \in \mathcal{X}} \frac{P(x) \, Q(x)}{R_\lambda(x)}
\leq \frac{1}{1-\lambda} \sum_{x \in \mathcal{X}} Q(x)
=\frac{1}{1-\lambda},\\[0.1cm]
\langle f_1,f_1 \rangle_\lambda
&=\lambda^2 \sum_{x \in \mathcal{X}} \frac{Q(x) \, \Delta(x)^2}{P(x) \, R_\lambda(x)}
\leq \lambda \sum_{x \in \mathcal{X}} \frac{\Delta(x)^2}{P(x)}
=\lambda B,\\[0.1cm]
\langle f_2,f_2\rangle_\lambda
&=(1-\lambda)^2 \sum_{x \in \mathcal{X}} \frac{P(x) \, \Delta(x)^2}{Q(x) \, R_\lambda(x)}
\leq(1-\lambda) \sum_{x \in \mathcal{X}} \frac{\Delta(x)^2}{Q(x)}
=(1-\lambda)A.
\end{align}
\end{subequations}
By the Cauchy--Schwarz inequality, the inner products of any two
retained functions are absolutely convergent.

We now compute the entries of their Gram matrix. Since
\begin{align}
\label{eq3: 23.11.26}
P=R_\lambda+\lambda\Delta,\qquad
Q=R_\lambda-(1-\lambda)\Delta,
\end{align}
multiplication gives
\[
\frac{PQ}{R_\lambda}
=R_\lambda+(2\lambda-1)\Delta-t \, \frac{\Delta^2}{R_\lambda}.
\]
Using \(\underset{x \in \mathcal{X}}{\sum} R_\lambda(x)=1\) and \(\underset{x \in \mathcal{X}}{\sum} \Delta(x)=0\), 
we obtain 
\begin{align}
\label{eq4: 23.11.26}
\langle f_0,f_0\rangle_\lambda=1-tT_\lambda.
\end{align}
For the off-diagonal entries, whenever the corresponding functions
are retained, \eqref{eq0: 23.11.26}, 
\eqref{eq1: 23.11.26}, \eqref{eq2: 23.11.26}, and \eqref{eq3: 23.11.26} yield 
\begin{subequations}\label{eq:weighted-ןinner-products}
\begin{align}
\langle f_0,f_1\rangle_\lambda
&=\lambda \sum_{x \in \mathcal{X}} \frac{Q(x) \, \Delta(x)}{R_\lambda(x)}
 =\lambda \sum_{x \in \mathcal{X}} \left(\Delta(x)-(1-\lambda) \, 
                         \frac{\Delta(x)^2}{R_\lambda(x)}\right)
 =-tT_\lambda,\\
\langle f_0,f_2\rangle_\lambda
&=-(1-\lambda) \sum_{x \in \mathcal{X}} \frac{P(x) \, \Delta(x)}{R_\lambda(x)}
 =-(1-\lambda) \sum_{x \in \mathcal{X}} \left(\Delta(x)+\lambda \, 
                         \frac{\Delta(x)^2}{R_\lambda(x)}\right)
 =-tT_\lambda,\\
\langle f_1,f_2\rangle_\lambda
&=-\lambda(1-\lambda) \sum_{x \in \mathcal{X}}
  \frac{P(x) \, Q(x)}{R_\lambda(x)} \, \frac{\Delta(x)}{P(x)} \, \frac{\Delta(x)}{Q(x)}
 =-t \sum_{x \in \mathcal{X}} \frac{\Delta(x)^2}{R_\lambda(x)} 
 =-tT_\lambda.
\end{align}
\end{subequations}
For the remaining diagonal entries, observe that
\[
\frac{\lambda Q}{P R_\lambda}
=\frac{R_\lambda-(1-\lambda)P}{P R_\lambda}
=\frac{1}{P}-\frac{1-\lambda}{R_\lambda},
\qquad
\frac{(1-\lambda)P}{Q R_\lambda}
=\frac{R_\lambda-\lambda Q}{Q R_\lambda}
=\frac{1}{Q}-\frac{\lambda}{R_\lambda}.
\]
Consequently, if \(B<\infty\), then
\begin{align}
\label{eq5: 23.11.26}
\langle f_1,f_1\rangle_\lambda
=\lambda \sum_{x \in \mathcal{X}} \Delta(x)^2 \, 
  \left(\frac{1}{P(x)}-\frac{1-\lambda}{R_\lambda(x)}\right)
=\lambda B-tT_\lambda.
\end{align}
Similarly, if \(A<\infty\), then
\begin{align}
\label{eq6: 23.11.26}
\langle f_2,f_2\rangle_\lambda
=(1-\lambda) \sum_{x \in \mathcal{X}} \Delta(x)^2 \, 
  \left(\frac{1}{Q(x)}-\frac{\lambda}{R_\lambda(x)}\right)
=(1-\lambda)A-tT_\lambda.
\end{align}

Let \( \mathbf{G} \) be the Gram matrix of the retained functions, in the order
\(f_0,f_1,f_2\), with any omitted function skipped. Let \( \mathbf{D} \) be the
diagonal matrix with entry \(1\) for \(f_0\), entry \(\lambda B\)
for \(f_1\) if retained, and entry \((1-\lambda)A\) for \(f_2\) if
retained. The preceding calculations in \eqref{eq4: 23.11.26}--\eqref{eq6: 23.11.26} give
\[
\mathbf{G}=\mathbf{D}-tT_\lambda\,\mathbf 1\mathbf 1^{\mathsf T},
\]
where \(\mathbf 1\) is the column vector of ones of the corresponding
dimension.

Since \(P\ne Q\), every diagonal entry of \( \mathbf{D} \) is strictly positive.
Set \( \underline{c}=\mathbf{D}^{-1}\mathbf 1\) and \(s=\mathbf 1^{\mathsf T} \underline{c} \).
Since \(\mathbf{D}^{-1}\) has diagonal entries equal to the reciprocals
of those of \(\mathbf{D}\), the component of \( \underline{c} \) corresponding to
\(f_0\) is \(1\). If \(f_1\) is retained, its component is
\(1/(\lambda B)\), and if \(f_2\) is retained, its component is
\(1/\bigl((1-\lambda)A\bigr)\). Thus \(s=\mathbf{1}^{\mathsf T} \underline{c} \)
is the sum of the components corresponding to the retained functions.
With the convention \(1/\infty=0\), this gives
\[
s=1+\frac{1}{\lambda B}+\frac{1}{(1-\lambda)A}.
\]
Since \(\mathbf{D} \underline{c}=\mathbf 1\), we have
\( \underline{c}^{\mathsf T} \mathbf{D} \, \underline{c} 
= \underline{c}^{\mathsf T} \mathbf 1=s\).
The positive semidefiniteness of \( \mathbf{G} \) therefore gives
\[
0\leq \underline{c}^{\mathsf T} \mathbf{G} \, \underline{c}
=\underline{c}^{\mathsf T} \mathbf{D} \underline{c}-t T_\lambda \, 
(\underline{c}^{\mathsf T} \mathbf 1)^2
=s-tT_\lambda s^2.
\]
Dividing by \(ts^2>0\) and substituting \(t=\lambda(1-\lambda)\)
(see \eqref{eq0: 23.11.26}) yields
\[
T_\lambda\leq\frac{1}{ts}
=\left(\frac{1-\lambda}{B}+\frac{\lambda}{A}
       +\lambda(1-\lambda)\right)^{-1}.
\]
This is the bound in the form \eqref{eq:extended}. When \(A,B<\infty\),
multiplying the numerator and denominator by \(AB\) gives
the right-hand side of \eqref{eq:bound}.
\end{proof}

\begin{lemma}\label{lem:integral}
Let \(A,B>0\). Then, the function \( F \) defined in \eqref{eq: function F} 
has the integral representation 
\begin{align}
\label{eq:closed-form}
\frac{F(A,B)}{\log e} = 
\int_0^1 \lambda\cdot \frac{AB}{(1-\lambda)A+\lambda B+\lambda(1-\lambda)AB} \; \mathrm d\lambda.
\end{align}
\end{lemma}

\begin{proof}
The denominator in the integrand is strictly positive on
\([0,1]\), since
\[
(1-\lambda)A+\lambda B+\lambda(1-\lambda)AB
\geq \min\{A,B\}>0.
\]
Differentiating the logarithm of the denominator of the integrand in 
\eqref{eq:closed-form} gives
\begin{align*}
&\frac{\lambda AB}
{(1-\lambda)A+\lambda B+\lambda(1-\lambda)AB}\\
&\quad=
\frac{AB+B-A}{2}\,
\frac{1}{(1-\lambda)A+\lambda B+\lambda(1-\lambda)AB}\\
&\qquad\quad
-\frac12 \; \frac{\mathrm d}{\mathrm d\lambda} \bigl\{
\ln\bigl((1-\lambda)A+\lambda B+\lambda(1-\lambda)AB\bigr) \bigr\},
\end{align*}
so integration on both sides over \([0,1]\) gives
\begin{align}
&\int_0^1\frac{\lambda AB}
{(1-\lambda)A+\lambda B+\lambda(1-\lambda)AB}\,
\mathrm d\lambda \notag\\
&\quad=\frac12\ln\frac AB
+\frac{AB+B-A}{2}
\int_0^1
\frac{\mathrm d\lambda}
{(1-\lambda)A+\lambda B+\lambda(1-\lambda)AB}.
\label{eq:reduction}
\end{align}
To evaluate the remaining integral, set
\[
S:=A+B+AB, \qquad \Delta:=\sqrt{S^2-4AB}.
\]
Since \(S>2\sqrt{AB}\), we have \(0<\Delta<S\).
The substitution \(u=\lambda/(1-\lambda)\) gives
\begin{align*}
\int_0^1
\frac{\mathrm d\lambda}
     {(1-\lambda)A+\lambda B+\lambda(1-\lambda)AB}
&=\int_0^\infty \frac{\mathrm du}{Bu^2+Su+A} \\[0.1cm]
&=\left[ \frac{1}{\Delta} \ln \left( \frac{2Bu+S-\Delta}{2Bu+S+\Delta} \right)
\right]_{0}^{\infty} \\[0.1cm]
&=\frac{1}{\Delta}\ln \left(\frac{S+\Delta}{S-\Delta} \right) \\[0.1cm]
&=\frac{2}{\Delta}\ln \left( \frac{S+\Delta}{2\sqrt{AB}} \right) \\[0.1cm]
&= \frac{F(A,B)}{\log e},
\end{align*}
where the penultimate equality follows from the identity
\((S-\Delta)(S+\Delta)=4AB\), and the last equality holds 
by \eqref{eq: function F}.
\end{proof}

\begin{lemma}
\label{lemma: inequality A,B}
Let \(A,B>0\). Then, 
\begin{align}
\label{eq:claim}
F(A,B) < \log(1+A) -\frac{3A^2 \log e}{2\bigl[(1+A)^2(1+B)-1\bigr]}.
\end{align}
\end{lemma}

\begin{proof}
Set \(s=1-\lambda\) in \eqref{eq:closed-form}. Subtracting the resulting
integral from
\[
\frac{\log(1+A)}{\log e}=\int_0^1\frac{A}{1+As}\,\mathrm ds
\]
and simplifying gives
\[
\frac{\log(1+A)-F(A,B)}{A^2\log e}
=\int_0^1\frac{s}{(1+As)\bigl[As+B(1-s)(1+As)\bigr]}\,\mathrm ds.
\]
The substitution \(x=(1+A)s/(1+As)\) maps \([0,1]\) onto itself and yields
\[
\frac{\log(1+A)-F(A,B)}{A^2\log e}
=\int_0^1\frac{x}{u(x)}\,\mathrm dx,
\]
where 
\[
u(x)=Ax(1+A-Ax)+(1+A)^2B(1-x).
\]
Since \(u(x)>0\) on \([0,1]\), the Cauchy--Schwarz inequality gives
\[
\int_0^1\frac{x}{u(x)}\,\mathrm dx
\geq\frac{\bigl(\int_0^1 x\,\mathrm dx\bigr)^2}
{\int_0^1 xu(x)\,\mathrm dx}.
\]
The denominator satisfies
\begin{align*}
\int_0^1 xu(x)\,\mathrm dx
&=\frac{A(1+A)}{3}-\frac{A^2}{4}+\frac{(1+A)^2B}{6}\\
&=\frac{(1+A)^2(1+B)-1-\tfrac12 A^2}{6}.
\end{align*}
Consequently,
\begin{align*}
\frac{\log(1+A)-F(A,B)}{A^2\log e}
&\geq \frac{3}{2\bigl[(1+A)^2(1+B)-1-\tfrac12 A^2\bigr]}\\
&> \frac{3}{2\bigl[(1+A)^2(1+B)-1\bigr]},
\end{align*}
where the strict inequality follows from \(A>0\). Rearranging proves
\eqref{eq:claim}.
\end{proof}

\begin{lemma}[Small-\(B\) asymptotics]
\label{lemma: small-B asymptotics}
Let \(F\) be given by \eqref{eq: function F}, and define
\[
G(A,B)=\log(1+A)
 -\frac{3A^2\log e}{2\bigl((1+A)^2(1+B)-1\bigr)},
\qquad A,B>0.
\]
For each fixed \(A>0\), let
\[
g(A)=\log(1+A)-\frac{3A\log e}{2(A+2)}.
\]
Then \(g(A)>0\), and as \(B\to0^+\),
\begin{align}
F(A,B)
 &= B\log\frac{A}{eB}
    +\mathrm{O}\left(B^2\log\frac{1}{B}\right),
 \label{eq:Ffirst}\\
G(A,B)
 &= g(A)+\frac{3(A+1)^2\log e}{2(A+2)^2}\,B
    +\mathrm{O}(B^2).
 \label{eq:Gfirst}
\end{align}
The implicit constants may depend on \(A\) and on the logarithm base.

For each fixed \(A>0\),
\begin{gather}
\lim_{B \to 0^+} F(A,B) = 0,
\qquad
\lim_{B \to 0^+} G(A,B) = g(A),
\label{eq:inner}\\
\frac{G(A,B)}{F(A,B)} \sim \frac{g(A)}{B \log(1/B)}
\longrightarrow +\infty \qquad (B \to 0^+).
\label{eq:ratio}
\end{gather}
The iterated limits are
\begin{equation}
\lim_{A \to \infty} \lim_{B \to 0^+} F(A,B) = 0,
\qquad
\lim_{A \to \infty} \lim_{B \to 0^+} G(A,B) = +\infty.
\label{eq:iterated}
\end{equation}
\end{lemma}

\begin{proof}
Fix \(A>0\), and let \(r:=\frac{B}{A}\) and \(d_\lambda:=1-\lambda+r\lambda\) for \( \lambda \in [0,1] \).
By Lemma~\ref{lem:integral},
\[
\frac{F(A,B)}{B\log e}
=\int_0^1\frac{\lambda}{d_\lambda+B\lambda(1-\lambda)} \; \mathrm{d}\lambda.
\]
Since \(0\leq\lambda(1-\lambda)\leq d_\lambda\) on \([0,1]\),
the denominator lies between \(d_\lambda\) and \((1+B)d_\lambda\).
Thus, as \(B \to 0^+\),
\[
\frac{F(A,B)}{B\log e}
=\bigl(1+\mathrm{O}(B)\bigr)
 \int_0^1\frac{\lambda}{1-(1-r)\lambda}\,\mathrm{d}\lambda.
\]
For \(0<r<1\), the last integral equals
\[
\frac{\ln \tfrac1r -1+r}{(1-r)^2}
=\ln \tfrac{1}{r}-1+\mathrm{O}\bigl(r\ln \tfrac1r \bigr).
\]
Substituting \(r=\frac{B}{A}\) gives \eqref{eq:Ffirst}.

For \eqref{eq:Gfirst}, we write the denominator in \(G\) as
\(A(A+2)+(A+1)^2B\). A first-order Taylor expansion at \(B=0\) gives
\[
G(A,B)=g(A)+\frac{3(A+1)^2\log e}{2(A+2)^2}\,B
       +\mathrm{O}(B^2).
\]
Moreover, \(g(0)=0\) by continuity, and
\[
g'(A)=\frac{(A^2+A+1)\log e}{(1+A)(A+2)^2}>0,
\]
so \(g(A)>0\) for every \(A>0\).

Since \(F(A,B)\sim B\log(1/B)\) and \(G(A,B)\to g(A)>0\) as
\(B\to0^+\), \eqref{eq:inner} and \eqref{eq:ratio} follow.
Finally, \(g(A)\to+\infty\) as \(A\to\infty\), which gives
\eqref{eq:iterated}.
\end{proof}

We finally prove Proposition~\ref{proposition: KL and chi^2} by relying 
on Lemmata~\ref{lemma: UB by chi squares}--\ref{lemma: small-B asymptotics}.
\begin{proof}
We start by proving \eqref{eq1: 11.11.26}.
For \(\lambda\in[0,1]\), define the probability mass function 
\[
R_\lambda(x):=(1-\lambda) P(x) + \lambda Q(x), \quad x \in \mathcal{X}.
\]
Then, 
\begin{align}
& \hspace*{-0.2cm} \tfrac1{\log e} \; D(P\|Q) \nonumber \\
&= \int_0^1 \chi^2(P \| R_\lambda) \; \mathrm{d}\lambda \nonumber \\
&= \int_0^1 \frac1{\lambda} \sum_{x \in \mathcal{X}} \frac{\bigl(P(x)-R_\lambda(x)\bigr)^2}{R_\lambda(x)} \; \mathrm{d}\lambda \nonumber \\
\label{eq3: 11.11.26}
&\leq \int_0^1 \lambda \cdot \frac{AB}{(1-\lambda)A+\lambda B+\lambda(1-\lambda)AB} \; \mathrm{d}\lambda \nonumber \\
&= \frac{F(A,B)}{\log e},
\end{align}
where the first equality is Eq.~(24) of \cite{NishiyamaS20}, the last inequality is due to Lemma~\ref{lemma: UB by chi squares}, 
and the last equality holds by Lemma~\ref{lem:integral}.

We next obtain the loosened upper bound on \( D(P \| Q) \) in \eqref{eq2: 11.11.26}. Let $A, B > 0$. 
By \eqref{eq3: 11.11.26} and Jensen's inequality, 
\begin{align}
\tfrac1{\log e} \; F(A,B) & = \int_0^1 \lambda \cdot \frac{AB}{(1-\lambda)A+\lambda B+\lambda(1-\lambda)AB} \; \mathrm{d}\lambda \\
& < \int_0^1 \lambda \cdot \frac{AB}{(1-\lambda)A+\lambda B} \; \mathrm{d}\lambda \\
& \leq AB \int_0^1 \lambda \left( \frac{1-\lambda}{A}+\frac{\lambda}{B} \right) \mathrm{d}\lambda \\
&= B\int_0^1\lambda(1-\lambda)\,\mathrm d\lambda +A\int_0^1\lambda^2\,\mathrm d\lambda \\[0.1cm]
\label{eq4: 11.11.26}
&=\tfrac13 A+\tfrac16 B.
\end{align}

By Lemma~\ref{lemma: inequality A,B}, it follows that the upper bound $F(A,B)$ 
on the right-hand side of \eqref{eq1: 11.11.26} strictly improves the upper bound 
\[
D(P\|Q) \leq \log\bigl(1+A\bigr),
\]
where \( A := \chi^2(P\|Q)\).
Furthermore, for fixed \(A>0\), the integrand increases with \(B\)
and converges to \( \frac{A}{1+A(1-u)}\) for \(0<u<1\).
By the monotone convergence theorem,
\[
\sup_{B>0}F(A,B)
=\lim_{B\to\infty}F(A,B)
=(\log e) \, \int_0^1\frac{A}{1+A(1-u)}\, \mathrm{d}u
=\log(1+A).
\]
Hence \(\log(1+A)\) is the optimal upper bound depending only on \(A\),
although it is not attained for any finite \(B>0\). This completes the 
proof of Item~1 of Proposition~\ref{proposition: KL and chi^2}.

Items~2 and~3 of Proposition~\ref{proposition: KL and chi^2} hold by 
Lemma~\ref{lemma: inequality A,B} and Lemma~\ref{lemma: small-B asymptotics}, 
respectively. 

We finally prove Item~4 of Proposition~\ref{proposition: KL and chi^2}.
Let \(\{P_n\}\) be a sequence of probability mass functions on a finite
or countably infinite alphabet \(\mathcal X\), and let \(Q\) be a
probability mass function on \(\mathcal X\). Assume that \(P_n\ll Q\)
for all sufficiently large \(n\) and equality \eqref{eq: uniform convergence of LR}
holds. Define
\[
A_n := \chi^2(P_n\|Q),
\qquad
B_n := \chi^2(Q\|P_n),
\]
and assume that \(0<A_n,B_n<\infty\) for all sufficiently large \(n\).
These assumptions ensure that \(P_n\) and \(Q\) have the same support
for all sufficiently large \(n\).

For an \(f\)-divergence whose convex generating function \(f\) satisfies
\(f(1)=0\) and is twice continuously differentiable in a neighborhood
of~\(1\), we have
\begin{equation}
\label{eq1: 13.11.26}
\lim_{n\to\infty}
\frac{D_f(P_n\|Q)}{\chi^2(P_n\|Q)}
= \tfrac12\,f''(1).
\end{equation}
Applying \eqref{eq1: 13.11.26} to the convex functions
\(f_1,f_2\colon(0,\infty)\to\mathbb{R}\), defined by
\(f_1(t)=t\log t\) and \(f_2(t)=(t-1)^2/t\), respectively, yields
\begin{align}
\lim_{n\to\infty}
\frac{D(P_n\|Q)}{A_n}
&= \tfrac12\,f_1''(1)
 = \tfrac12\,\log e,\\
\lim_{n\to\infty}
\frac{B_n}{A_n}
&= \tfrac12\,f_2''(1)
 = 1.
\end{align}
Consequently, the upper bounds on \(D(P_n\|Q)\) on the right-hand sides of 
\eqref{eq1: 11.11.26} and \eqref{eq2: 11.11.26} are locally tight:
the ratio of each bound to \(D(P_n\|Q)\) tends to~\(1\) as
\(n\to\infty\). This establishes \eqref{eq2: 13.11.26} and completes
the proof of Proposition~\ref{proposition: KL and chi^2}. 
\end{proof}

\begin{remark}
{\em On a countably infinite alphabet, pointwise convergence of \(P_n\)
to \(Q\) does not suffice for the local quadratic approximation to 
yield \eqref{eq1: 13.11.26}, 
even if both \(\chi^2(P_n\|Q)\) and \(\chi^2(Q\|P_n)\) tend to zero.
Thus, the uniform convergence of the likelihood ratios cannot, in
general, be replaced by pointwise convergence alone.

To see this, let \(\mathcal X=\mathbb N\) and \(Q(k)=2^{-k}\) for 
all \( k \in \mathbb{N} \).
For \(n\ge2\), let \(\varepsilon_n=2^{-n}\) and define
\[
P_n(k)=
\begin{cases}
\frac12-\varepsilon_n, & k=1,\\
2\varepsilon_n, & k=n,\\
2^{-k}, & k\notin\{1,n\}.
\end{cases}
\]
Then \(P_n\) and \(Q\) are probability mass functions that 
have the same support, and 
\( P_n(k)\underset{n\to\infty}{\longrightarrow}Q(k) \)
for every fixed \(k \in \mathbb N\). Furthermore,
\[
\chi^2(P_n\|Q)
=\varepsilon_n+2\varepsilon_n^2 \longrightarrow 0,
\qquad
\chi^2(Q\|P_n)
=\frac{\varepsilon_n}{2}
+\frac{\varepsilon_n^2}{\frac12-\varepsilon_n}
\longrightarrow 0.
\]
Nevertheless, for \(f(t)=t\log t\) \((t>0)\), 
\[
\begin{aligned}
D_f(P_n\|Q) &= D(P_n \| Q) \\
&=\left(\tfrac12-\varepsilon_n\right) \log(1-2\varepsilon_n)
   +2 \varepsilon_n \log2 \\
&=(2 \log 2-\log e) \varepsilon_n + O(\varepsilon_n^2).
\end{aligned}
\]
Consequently,
\begin{align*}
\lim_{n \to \infty} 
\frac{D_f(P_n\|Q)}{\chi^2(P_n\|Q)}
&=2 \log 2 - \log e \\
&\ne \tfrac12 \, \log e 
=\tfrac12 f''(1).
\end{align*}
Here the likelihood ratios converge pointwise to \(1\), but
\[
\sup_{k \geq 1}
\left|\frac{P_n(k)}{Q(k)}-1\right|=1
\qquad(n \geq 2),
\]
since \( \frac{P_n(n)}{Q(n)}=2\) for every \( n \in \mathbb{N} \). The 
failure therefore arises from a
nonvanishing relative perturbation at a symbol whose \(Q\)-probability
tends to zero.}
\end{remark}

\subsection{Maximum relative entropy between binomial distributions under chi-squared divergence constraints and its properties}
\label{subsection: KL between binomial distributions}

Let $\mathbb{N}=\{1,2,\ldots\}$ denote the set of natural numbers.
For $n\in\mathbb{N}$ and $p,q\in(0,1)$, let $P$ and $Q$ denote
the probability mass functions of the binomial distributions with
parameters $(n,p)$ and $(n,q)$, respectively:
\[
P(k)=\binom{n}{k}p^k(1-p)^{n-k},
\qquad
Q(k)=\binom{n}{k}q^k(1-q)^{n-k},
\qquad k\in\{0,1,\ldots,n\}.
\]
Let 
\[
d(p\|q) = p \log \frac{p}{q} + (1-p) \log \frac{1-p}{1-q}
\]
denote the binary relative entropy, i.e., the relative entropy
from the Bernoulli distribution with probability vector $(p,1-p)$
to that with probability vector $(q,1-q)$, or equivalently,
from $(1-p,p)$ to $(1-q,q)$. A direct calculation gives
\begin{align}
\label{eq1: binomial}
& \chi^2(P\|Q) = \left( \frac{p^2}{q} + \frac{(1-p)^2}{1-q} \right)^n - 1, \quad 
& D(P\|Q) = n d(p\|q),  \\
\label{eq2: binomial}
& \chi^2(Q\|P) = \left( \frac{q^2}{p} + \frac{(1-q)^2}{1-p} \right)^n - 1, \quad 
& D(Q\|P) = n d(q\|p).
\end{align}

\begin{proposition}[Binomial distributions with prescribed ordered chi-squared divergences]
\label{prop:binomial-prescribed-divergences}
Fix $A,B>0$. For each $n\in\mathbb{N}$, let
\begin{equation}
\label{eq1: parameter-definitions}
a_n := (1+A)^{1/n}-1, \qquad b_n := (1+B)^{1/n}-1,
\end{equation}
and define
\begin{subequations}
\label{eq2: parameter-definitions}
\begin{align}
\label{eq: R_n}
R_n&:= \sqrt{(a_n-b_n)^2+2a_n b_n (a_n+b_n) +a_n^2 b_n^2},\\[0.1cm]
\label{eq: u_n, v_n}
u_n&:= \frac{a_n b_n - a_n + b_n}{R_n},\qquad
v_n:= \frac{a_n b_n + a_n-b_n}{R_n},\\[0.1cm]
\label{eq: p_n, q_n}
p_n&:= \frac{1+u_n}{2},\qquad
q_n:= \frac{1-v_n}{2}.
\end{align}
\end{subequations}
For $p,q\in(0,1)$, let $P_n$ and $Q_n$ denote the probability mass
functions of $\operatorname{Bin}(n,p)$ and $\operatorname{Bin}(n,q)$,
respectively. The constraints
\begin{equation}\label{eq:prescribed-chi-square}
\chi^2(P_n\|Q_n)=A, \qquad \chi^2(Q_n\|P_n)=B
\end{equation}
hold for exactly two parameter pairs in $(0,1)^2$, namely
\begin{equation}\label{eq:parameter-pairs}
(p,q)=(p_n,q_n) \quad \text{and} \quad (p,q)=(1-p_n,1-q_n).
\end{equation}
The first pair is the unique solution with $p>q$.

For each $n\in\mathbb{N}$, either choice in \eqref{eq:parameter-pairs}
gives the same values of the two ordered relative entropies:
\begin{subequations}
\begin{align}
\label{eq:forward-KL}
D(P_n\|Q_n) &= n\,d(p_n\|q_n), \\
\label{eq:reverse-KL}
D(Q_n\|P_n) &= n\,d(q_n\|p_n).
\end{align}
\end{subequations}
With $A$ and $B$ fixed, both sequences are strictly decreasing in $n$.
Hence their maxima are attained simultaneously and only at $n=1$, with
\begin{subequations}
\begin{align}
\label{eq:maximum-forward}
\max_{n \in \mathbb{N}} D(P_n\|Q_n) &= d(p_1\|q_1), \\
\label{eq:maximum-reverse}
\max_{n \in \mathbb{N}} D(Q_n\|P_n) &= d(q_1\|p_1). 
\end{align}
\end{subequations}
Each of the Bernoulli parameter pairs $(p_1,q_1)$ and $(1-p_1,1-q_1)$
attains both maxima, where
\begin{equation}\label{eq1: 14.11.26}
p_1 = \frac12 \left(1 + \frac{AB-A+B}{R_1}\right),\qquad
q_1 = \frac12 \left(1 - \frac{AB+A-B}{R_1}\right),
\end{equation}
and
\begin{equation}\label{eq2: 14.11.26}
R_1 = \sqrt{(A-B)^2 + 2AB (A+B) + A^2 B^2}.
\end{equation}

Let $\alpha:=\log(1+A)$ and $\beta:=\log(1+B)$, and define
\begin{equation}
\label{eq:binomial-limit-coefficient}
L(A,B):=
\begin{cases}
\displaystyle \frac{\alpha\beta}{(\alpha-\beta)^2}
\left(\frac{\alpha}{\log e}\log\frac{\alpha}{\beta}-\alpha+\beta\right),
& A\ne B,\\[6pt]
\displaystyle \tfrac12\log(1+A), & A=B.
\end{cases}
\end{equation}
The two relative entropies converge from above to their respective limits:
\begin{subequations}
\begin{align}
\label{eq:forward-limit}
\lim_{n \to \infty} D(P_n\|Q_n) &= L(A,B), \\
\label{eq:reverse-limit}
\lim_{n \to \infty} D(Q_n\|P_n) &= L(B,A).
\end{align}
\end{subequations}
For fixed $A,B>0$, the errors satisfy
\begin{subequations}
\begin{align}
\label{eq:forward-rate}
D(P_n\|Q_n)-L(A,B) &= O(n^{-1}), \\
\label{eq:reverse-rate}
D(Q_n\|P_n)-L(B,A) &= O(n^{-1}).
\end{align}
\end{subequations}
Both limits are strictly positive, and their off-diagonal formulas extend
continuously to the common value $\tfrac12\log(1+A)$ at $A=B$.
For every $n\in\mathbb{N}$,
\begin{subequations}
\label{eq: decreasing KL}
\begin{align}
\label{eq1: decreasing KL}
0<L(A,B)&< D(P_{n+1}\|Q_{n+1}) < D(P_n\|Q_n),\\
\label{eq2: decreasing KL}
0<L(B,A)&< D(Q_{n+1}\|P_{n+1}) < D(Q_n\|P_n).
\end{align}
\end{subequations}
\end{proposition}

\begin{proof}
We first reduce the two prescribed chi-squared constraints to equations
for $p$ and $q$ and show that their solutions form a complementary pair
of parameter choices. We then prove strict monotonicity by viewing
$D(P_n\|Q_n)$ as the value at $t=\frac1n$ of a function whose derivative
is positive by Lemma~\ref{lem:binary-relative-entropy-differential-inequality}.
Interchanging $A$ and $B$ gives the reverse direction, so both maxima
occur at $n=1$. Finally, we derive an integral representation that gives
the limiting values and the $O(n^{-1})$ convergence rates in both cases
$A=B$ and $A\ne B$.

\medskip\noindent
\emph{Parameter pairs and relative entropies.}
Fix $n\in\mathbb{N}$ and set $\delta:=p-q$. By
\eqref{eq1: binomial}--\eqref{eq2: binomial} and the identity
\[
\frac{p^2}{q}+\frac{(1-p)^2}{1-q}
=1+\frac{(p-q)^2}{q(1-q)},
\]
the constraints \eqref{eq:prescribed-chi-square} are equivalent to
\begin{equation}
\label{eq:reduced-system}
\delta^2=a_n q(1-q)=b_n p(1-p).
\end{equation}
Since $a_n, b_n > 0$ (see \eqref{eq1: parameter-definitions}, where $A,B>0$
are fixed) and $p,q \in (0,1)$, we have $\delta \neq 0$. By \eqref{eq:reduced-system}, 
\[
\delta(1-p-q)
=p(1-p)-q(1-q)
=\delta^2\left(\frac1{b_n}-\frac1{a_n}\right).
\]
Dividing by $\delta$ and using $p-q=\delta$, we obtain
\begin{align}
\label{eq: 2p-1, 2q-1}
2p-1 = \left( \frac{a_nb_n-a_n+b_n}{a_nb_n} \right) \delta, \qquad
1-2q = \left( \frac{a_nb_n+a_n-b_n}{a_nb_n} \right) \delta.
\end{align}
By \eqref{eq:reduced-system}, we get 
\(
4\delta^2 = 4a_nq(1-q) = a_n\bigl[1-(1-2q)^2\bigr].
\)
Substituting the last expression for \(1-2q\) and solving for \(\delta^2\) yields
\[
\delta^2
=\frac{a_n^2 b_n^2}{(a_n b_n + a_n - b_n)^2+4 a_n b_n^2}
=\frac{a_n^2 b_n^2}{R_n^2} \quad \Longrightarrow \quad \delta=\pm \frac{a_n b_n}{R_n}.
\]
Substitution of the last equality for $\delta$ into the formulas for $2p-1$ and $1-2q$ 
in \eqref{eq: 2p-1, 2q-1} gives $(p_n,q_n)$ in \eqref{eq: p_n, q_n} for the positive sign 
and $(1-p_n,1-q_n)$ for the negative sign. To verify that both pairs are admissible, 
observe that by \eqref{eq: R_n}
\[
R_n^2-(a_n b_n-a_n+b_n)^2=4a_n^2 b_n,\qquad
R_n^2-(a_n b_n+a_n-b_n)^2=4a_n b_n^2.
\]
Hence $|u_n|,|v_n|<1$, so $p_n,q_n\in(0,1)$ (see \eqref{eq: u_n, v_n} and \eqref{eq: p_n, q_n}). 
Moreover,
\[
p_n(1-p_n)=\frac{1-u_n^2}{4}=\frac{a_n^2 b_n}{R_n^2},
\qquad
q_n(1-q_n)=\frac{1-v_n^2}{4}=\frac{a_n b_n^2}{R_n^2}.
\]
Since $p_n-q_n = \frac{a_n b_n}{R_n}>0$, these identities verify
\eqref{eq:reduced-system} for $(p_n,q_n)$. Taking complements
preserves both variances and reverses the sign of $p_n-q_n$, so
$(1-p_n,1-q_n)$ also satisfies \eqref{eq:reduced-system}.
This proves \eqref{eq:parameter-pairs}, with $(p_n,q_n)$ the unique
solution for which $p>q$.

By \eqref{eq1: binomial}--\eqref{eq2: binomial}, the ordered
relative entropies for the first pair are $n\,d(p_n\|q_n)$ and
$n\,d(q_n\|p_n)$. Since $d(1-p\|1-q)=d(p\|q)$, the complementary
pair gives the same two values. This proves
\eqref{eq:forward-KL}--\eqref{eq:reverse-KL}.

\medskip\noindent
\emph{Strict monotonicity and maxima.}
For $a,b>0$, let
\(
M(a,b):=d(p\|q),
\)
where $(p,q)$ is the unique solution with $p>q$ of
\[
(p-q)^2=a\,q(1-q)=b\,p(1-p).
\]
The existence and uniqueness of $(p,q)$ follow from the parameter calculation above
by setting $n=1$, $A=a$, and $B=b$ in \eqref{eq:reduced-system}. The explicit formulas 
also show that $p$ and $q$ depend differentiably on $a,b>0$, since the
radicand in their common denominator is strictly positive.
Interchanging $a$ and $b$ replaces $(p,q)$ by $(1-q,1-p)$, so
\[
M(b,a)=d(1-q\|1-p)=d(q\|p).
\]

For $t>0$, define
\[
a(t):=(1+A)^t-1, \qquad b(t):=(1+B)^t-1,\qquad
J(t):=\frac{M(a(t),b(t))}{t}.
\]
Then $a\bigl(\frac1n\bigr)=a_n$, $b\bigl(\frac1n\bigr)=b_n$, and
\[
J\bigl(\tfrac1n\bigr)=nM(a_n,b_n)=D(P_n\|Q_n).
\]
Differentiating $a(t)$ and $b(t)$ gives the identities 
\[
t\,a'(t)=(1+a(t))\ln(1+a(t)), \qquad
t\,b'(t)=(1+b(t))\ln(1+b(t)).
\]
By the chain rule,
\begin{align*}
t^2 J'(t)
={}&(1+a(t))\ln(1+a(t))\,M_a(a(t),b(t))\\
 &+(1+b(t))\ln(1+b(t))\,M_b(a(t),b(t))
   -M(a(t),b(t)),
\end{align*}
where $(M_a, M_b) := \bigl(\frac{\partial M}{\partial a}, \frac{\partial M}{\partial b} \bigr)$.
Lemma~\ref{lem:binary-relative-entropy-differential-inequality},
proved directly from the Bernoulli likelihood ratios in
Appendix~\ref{app:binary-relative-entropy}, gives
$t^2J'(t)>0$ for every $t>0$. Hence $J$ is
strictly increasing, and $D(P_n\|Q_n)=J\bigl(\frac1n\bigr)$ is strictly
decreasing in $n$. Interchanging $A$ and $B$ proves the same
assertion for $D(Q_n\|P_n)$. Both maxima are therefore attained
only at $n=1$, proving
\eqref{eq:maximum-forward}--\eqref{eq:maximum-reverse}.
Substituting $a_1=A$ and $b_1=B$ gives the Bernoulli parameters
in \eqref{eq1: 14.11.26}--\eqref{eq2: 14.11.26}.

\medskip\noindent
\emph{Limits and convergence rates.}
For fixed $q\in(0,1)$, differentiation gives
\[
\frac{\partial}{\partial x} \, d(x\|q)
=\log\frac{x(1-q)}{q(1-x)},\qquad
\frac{\partial^2}{\partial x^2} \, d(x\|q)
=\frac{\log e}{x(1-x)}.
\]
Both $d(x\|q)$ and its first derivative vanish at $x=q$.
Set $x_{n,s}:=(1-s)q_n+sp_n$ for $0\leq s\leq1$.
Taylor's formula with integral remainder gives
\[
d(p_n\|q_n)
=(\log e)(p_n-q_n)^2
\int_0^1\frac{1-s}{x_{n,s}(1-x_{n,s})}\,\mathrm{d}s.
\]
By \eqref{eq:reduced-system},
\begin{align*}
x_{n,s}(1-x_{n,s})
&=(1-s)q_n(1-q_n)+sp_n(1-p_n)+s(1-s)(p_n-q_n)^2\\
&=(p_n-q_n)^2
  \left(\frac{1-s}{a_n}+\frac{s}{b_n}+s(1-s)\right).
\end{align*}
Cancelling $(p_n-q_n)^2$ and multiplying by $n$, we obtain
\begin{equation}\label{eq:binomial-entropy-integral}
D(P_n\|Q_n)
=(\log e)\int_0^1
\frac{1-s}{(1-s)/(na_n)+s/(nb_n)+s(1-s)/n}\,\mathrm{d}s.
\end{equation}

Recall that $\alpha=\log(1+A)$ and $\beta=\log(1+B)$ in \eqref{eq:binomial-limit-coefficient}. Expanding
the exponential gives
\[
a_n=\exp\left(\frac{\alpha}{n\log e}\right)-1
=\frac{\alpha}{n\log e}+O(n^{-2}),\qquad
b_n=\frac{\beta}{n\log e}+O(n^{-2}).
\]
Since $\alpha,\beta>0$, it follows that
\[
\frac1{na_n}=\frac{\log e}{\alpha}+O(n^{-1}),\qquad
\frac1{nb_n}=\frac{\log e}{\beta}+O(n^{-1}).
\]
Thus the denominator in \eqref{eq:binomial-entropy-integral}
differs from $(\log e)((1-s)/\alpha+s/\beta)$ by $O(n^{-1})$
uniformly for $s\in[0,1]$. The latter expression is at least
$(\log e)/\max\{\alpha,\beta\}>0$, so its reciprocal also
approximates the reciprocal of the denominator with a uniform
$O(n^{-1})$ error. Consequently,
\[
D(P_n\|Q_n)
=\alpha\beta\int_0^1
\frac{1-s}{\beta+(\alpha-\beta)s}\,\mathrm{d}s+O(n^{-1}).
\]

If $\alpha\ne\beta$, the substitution $y=\beta+(\alpha-\beta)s$
gives
\begin{align*}
\alpha\beta\int_0^1\frac{1-s}{\beta+(\alpha-\beta)s}\,\mathrm{d}s
&=\frac{\alpha\beta}{(\alpha-\beta)^2}
  \int_\beta^\alpha\left(\frac{\alpha}{y}-1\right)\,\mathrm{d}y\\
&=\frac{\alpha\beta}{(\alpha-\beta)^2}
  \left(\frac{\alpha}{\log e}\log\frac{\alpha}{\beta}
        -\alpha+\beta\right).
\end{align*}
If $\alpha=\beta$, the same expression reduces to
$\alpha\int_0^1(1-s)\,\mathrm{d}s=\alpha/2$. Thus
\[
D(P_n\|Q_n)=L(A,B)+O(n^{-1}).
\]
Interchanging $A$ and $B$ gives
$D(Q_n\|P_n)=L(B,A)+O(n^{-1})$. This proves
\eqref{eq:forward-limit}--\eqref{eq:reverse-limit} and
\eqref{eq:forward-rate}--\eqref{eq:reverse-rate}.

Finally, the representation
\[
L(A,B)=\int_0^1
\frac{\alpha\beta(1-s)}{(1-s)\beta+s\alpha}\,\mathrm{d}s
\]
shows that $L(A,B)>0$, since the integrand is positive for $0\leq s<1$.
Its denominator is at least $\min\{\alpha,\beta\}$. Hence, on every
compact subset of $(0,\infty)^2$, the integrand depends continuously
on $(\alpha,\beta)$ uniformly in $s\in[0,1]$. The integral therefore
extends continuously to $\alpha=\beta$, where its value is
$\alpha/2=\tfrac12\log(1+A)$. The same conclusions hold for $L(B,A)$.

Since each sequence decreases strictly to its limit, inequalities 
\eqref{eq1: decreasing KL}--\eqref{eq2: decreasing KL} hold for every
$n \in \mathbb{N}$.
\end{proof}

\noindent 
\begin{table}[hbt]
    \centering
    \caption{\centering{Comparison of the ordered relative entropies where $P_n$ and $Q_n$ are the probability mass functions of
    $\operatorname{Bin}(n,p_n),\operatorname{Bin}(n,q_n)$, respectively, such that $\chi^2(P_n\|Q_n) = 10$
    and $\chi^2(Q_n\|P_n) = \tfrac1{10}$.}}
    \label{tab:KL-binomial}
    \vspace{10pt}
    \begin{tabular}{|c|c|c|}
        \hline
        & \multicolumn{2}{c|}{Relative entropies [nats]} \\
        \cline{2-3}
        \(n\) & \(D(P_n \| Q_n)\) & \(D(Q_n \| P_n)\) \\ \hline
        1 & $ 0.34727918 $ & $ 0.09211540 $ \\ \hline
        2 & $ 0.28560110 $ & $ 0.08953261 $ \\ \hline
        3 & $ 0.26721429 $ & $ 0.08844966 $ \\ \hline
        4 & $ 0.25847244 $ & $ 0.08787068 $ \\ \hline
        5 & $ 0.25337539 $ & $ 0.08751222 $ \\ \hline
       10 & $ 0.24351851 $ & $ 0.08677230 $ \\ \hline
   $10^2$ & $ 0.23503451 $ & $ 0.08608262 $ \\ \hline 
   $10^3$ & $ 0.23420633 $ & $ 0.08601253 $ \\ \hline
   $10^4$ & $ 0.23412372 $ & $ 0.08600551 $ \\ \hline 
   $10^5$ & $ 0.23411545 $ & $ 0.08600481 $ \\ \hline
   $10^6$ & $ 0.23411463 $ & $ 0.08600474 $ \\ \hline
    \end{tabular}
\end{table}
Numerical experimentation related to Proposition~\ref{prop:binomial-prescribed-divergences} 
is presented in Table~\ref{tab:KL-binomial}.
Both ordered relative entropies decrease from their Bernoulli maxima
toward the respective limits given by \eqref{eq:binomial-limit-coefficient},
namely $L\bigl(10,\tfrac1{10})\approx 0.23411454$ and
$L(\tfrac1{10},10)\approx 0.08600473$ nats. 
For large $n$, a tenfold increase in $n$ reduces each gap to its
limit by approximately a factor of ten.

\subsection{Attainability of the tight upper bound on the relative entropy under chi-squared divergence constraints}
\label{subsection: attainability}

\begin{proposition}
\label{proposition: attainability of UB}
Let $A, B > 0$ be arbitrarily fixed numbers. 
The upper bound on \( D(P\|Q) \) on the right-hand side of 
\eqref{eq1: 11.11.26} is attained by each of the two 
pairs of Bernoulli distributions 
\[
P = (p_1, 1-p_1),  \quad Q = (q_1, 1-q_1)
\]
and 
\[
P = (1-p_1, p_1),  \quad Q = (1-q_1, q_1),
\]
where $p_1, q_1 \in (0,1)$ are specified in 
Eqs.~\eqref{eq1: 14.11.26} and~\eqref{eq2: 14.11.26} of 
Proposition~\ref{prop:binomial-prescribed-divergences}.
Likewise, the upper bound on \( D( Q \| P) \), obtained by
interchanging \( A \) and \( B \) on the right-hand side of 
\eqref{eq1: 11.11.26}, is simultaneously attained 
by these two pairs of Bernoulli distributions.
\end{proposition}

\begin{proof}
Set $S=A+B+AB$ and $R=R_1$ (see \eqref{eq2: 14.11.26}). Expanding gives
\[
R^2=S^2-4AB.
\]
Since $S>2\sqrt{AB}$, we have $0<R<S$. Furthermore,
\[
R^2-(AB-A+B)^2=4A^2B>0,
\qquad
R^2-(AB+A-B)^2=4AB^2>0.
\]
Consequently, $p_1,q_1\in(0,1)$.
Direct substitution and simplification yield
\[
\frac{p_1}{q_1}=\frac{S+R}{2B},
\qquad
\frac{1-p_1}{1-q_1}=\frac{S-R}{2B}.
\]
Let 
\[
L=\log\left(\frac{S+R}{2\sqrt{AB}}\right).
\]
Using $(S+R)(S-R)=4AB$, we obtain
\[
\log\frac{p_1}{q_1}
=\frac12\log\frac AB+L,
\qquad
\log\frac{1-p_1}{1-q_1}
=\frac12\log\frac AB-L,
\]
and
{\small 
\begin{align*}
D(P\|Q) &= d(p_1\|q_1) \\
&=p_1\left(\frac12\log\frac AB+L\right)
 +(1-p_1)\left(\frac12\log\frac AB-L\right)\\
&=\frac12\log\frac AB+(2p_1-1)L\\
&=\frac12\log\frac AB+\frac{(AB+B-A)L}{R} \\
&=F(A,B),
\end{align*}}
which proves the first part of our assertion. Similarly,
{\small
\begin{align*}
D(Q\|P) &= d(q_1\|p_1) \\
&=-q_1\left(\frac12\log\frac AB+L\right)
  -(1-q_1)\left(\frac12\log\frac AB-L\right)\\
&=\frac12\log\frac BA+(1-2q_1)L\\
&=\frac12\log\frac BA+\frac{AB+A-B}{R}L \\
&=F(B,A),
\end{align*}}
which proves the second part.
\end{proof}

\subsection{Limits and small-parameter scaling of the tight relative entropy bound}
\label{subsection: asymptotic}

By Propositions~\ref{proposition: KL and chi^2}
and~\ref{proposition: attainability of UB}, the upper bound
$F(A,B)$ in \eqref{eq: function F} is attained by binary
distributions satisfying $\chi^2(P\|Q)=A$ and $\chi^2(Q\|P)=B$.
We determine how $F(A,B)$ scales as $A$ and $B$ tend to zero
at arbitrary relative rates. We also derive its expansion for
fixed $B$ as $A \to 0^+$ and its behavior as $B \to \infty$
with $A$ fixed or tending to zero.

\begin{proposition}[Limits and small-parameter scaling]
For $A,B>0$, define the continuous function
\[
 G(A,B)=\begin{cases}
 \displaystyle\frac{AB}{(B-A)^2}
 \left[(B-A)\log e-A\log\frac BA\right], &A \ne B,\\[7pt]
 \displaystyle \tfrac12 A \log e, &A=B.
 \end{cases}
\]
Then the following assertions hold.
\begin{enumerate}
\item As $A,B\to0^+$ jointly, with no restriction on their ratio,
\[
 F(A,B)=G(A,B)\bigl[1+O(\min\{A,B\})\bigr],
 \qquad \lim_{\substack{A\to0^+\\B\to0^+}} F(A,B)=0,
\]
where the relative error estimate is uniform in $A/B$.
In particular,
\[
 \begin{array}{c|c}
 \text{\normalfont Relative scaling}&\text{\normalfont Leading asymptotic}\\ \hline
 A/B\to0&F(A,B)\sim A\log e\\[2pt]
 A/B\to1&F(A,B)\sim \frac A2\log e\\[2pt]
 B/A\to0&F(A,B)\sim B\log(A/B)
 \end{array}
\]
More generally, if $B/A \to r\in(0,\infty)$, then
\[
 \frac{F(A,B)}{A} \longrightarrow
 \begin{cases}
 \displaystyle\frac{r[(r-1)\log e-\log r]}{(r-1)^2}, & r \ne 1,\\[6pt]
 \frac12\log e, & r=1.
 \end{cases}
\]
\item For every fixed $A > 0$,
\[
 \lim_{B\to\infty} F(A,B)=\log(1+A).
\]
\item For every fixed $B > 0$, as $A \to 0^+$,
\[
 F(A,B)=A\log e-\frac{A^2}{B} \log\frac{B}{A}
 +\left(\frac1B-\frac12 \right)A^2\log e
 +O_B\left(A^3 |\log A|\right).
\]
In particular, $F(A,B)\sim A\log e$ and $F(A,B)\to0$.
\item As $A\to0^+$ and $B\to\infty$ jointly,
\[
 F(A,B)\sim A\log e,
 \qquad \lim_{\substack{A \to 0^+ \\B \to \infty}} F(A,B)=0.
\]
\end{enumerate}
\end{proposition}

\begin{proof}
By Lemma~\ref{lem:integral}, the substitution $t=1-\lambda$ gives
\begin{equation}\label{eq:integral}
F(A,B)=AB\log e\int_0^1
\frac{1-t}{Q_0(t)+ABt(1-t)}\,\mathrm{d}t,
\qquad Q_0(t):=B(1-t)+At.
\end{equation}
Omitting the term $ABt(1-t)$ from the denominator gives, by direct
integration,
\[
AB\log e\int_0^1\frac{1-t}{Q_0(t)}\,\mathrm{d}t=G(A,B).
\]
Since $Q_0(t)\ge B(1-t)$ and $Q_0(t)\ge At$, we have
\[
0\le \frac{ABt(1-t)}{Q_0(t)}
\le \min\{At,B(1-t)\}\le m,
\qquad m:=\min\{A,B\}.
\]
It follows that
\[
Q_0(t)\le Q_0(t)+ABt(1-t)\le (1+m)Q_0(t).
\]
Taking reciprocals and multiplying by $AB(1-t)$ yields, for every
$t\in[0,1]$,
\[
\frac{1}{1+m}\,\frac{AB(1-t)}{Q_0(t)}
\le \frac{AB(1-t)}{Q_0(t)+ABt(1-t)}
\le \frac{AB(1-t)}{Q_0(t)}.
\]
The rightmost expression is at most $A$, since $B(1-t)\le Q_0(t)$.
Integrating over $0\le t\le1$ and multiplying by $\log e$ therefore
gives, for all $A,B>0$,
\begin{equation}\label{eq:bounds}
\frac{G(A,B)}{1+\min\{A,B\}}
\le F(A,B)\le G(A,B)\le A\log e.
\end{equation}
In particular,
\[
0\le 1-\frac{F(A,B)}{G(A,B)}
\le \frac{m}{1+m}\le m.
\]
This proves the relative error estimate in assertion~1 uniformly in
$A/B$. The joint zero limit follows from $0<F(A,B)\le A\log e$.

For the ratio-dependent limits, write $G(A,B)=A\phi(B/A)$, where
\[
\phi(r)=\frac{r[(r-1)\log e-\log r]}{(r-1)^2}
\quad (r \ne 1), \qquad \phi(1)=\tfrac12 \log e.
\]
The function $\phi$ is continuous at $r=1$, tends to $\log e$ as
$r\to\infty$, and satisfies $\phi(r)\sim r\log(1/r)$ as $r\to0^+$.
Together with \eqref{eq:bounds}, these facts give the remaining
claims in assertion~1. They also prove assertion~4. Indeed, if
$A\to0^+$ and $B\to\infty$, then $m\to0$ and $B/A\to\infty$.
Hence $F(A,B)/G(A,B) \to 1$ and $G(A,B)/A \to \log e$, so
$F(A,B)\sim A \log e \to 0$.

For assertion~2, fix $A>0$. For $0\le t<1$, the fraction in
\eqref{eq:integral}, including the prefactor $AB$, is bounded by
$A$ and converges to $A/(1+At)$ as $B\to\infty$. Hence, by
dominated convergence,
\[
\lim_{B\to\infty}F(A,B)
=A\log e\int_0^1\frac{\mathrm{d}t}{1+At}
=\log(1+A).
\]

For assertion~3, fix $B>0$ and set
\[
S:=A+B+AB,\qquad D:=\sqrt{S^2-4AB},\qquad N:=AB+B-A.
\]
Since $D^2-N^2=4A^2B$ and $D+N=2B+O_B(A)$ as $A \to 0^+$,
\[
D-N=\frac{4A^2B}{D+N}=2A^2+O_B(A^3),
\qquad
\frac ND-1=-\frac{2A^2}{B}+O_B(A^3).
\]
Also, $S+N=2B(1+A)$, so
\[
J:=\log\frac{S+D}{2B}
=\log\left(1+A+\frac{D-N}{2B}\right)
=\left[A+\left(\frac1B-\frac12\right)A^2\right]\log e
 +O_B(A^3).
\]
Rewriting \eqref{eq: function F} as
\[
F(A,B)=J+\left(\frac ND-1\right)
             \left(\frac12\log\frac BA+J\right)
\]
and substituting the preceding expansions gives
\[
F(A,B)=A\log e-\frac{A^2}{B}\log\frac BA
+\left(\frac1B-\frac12\right)A^2\log e
+O_B\bigl(A^3|\log A|\bigr).
\]
In particular, $F(A,B)\sim A\log e$ as $A \to 0^+$, completing
the proof of assertion~3.
\end{proof}

\section{Redundancy Reduction by Mixing in Lossless Source Coding}
\label{section: redundancy reduction by mixing in lossless source coding}

In lossless source coding, the ideal codelength assigned to a symbol
$x$ under a model $S$ is $-\log_2 S(x)$ bits~\cite[Ch.~5]{CoverThomas2006}.
Suppose that an encoder knows the laws $P$ and $Q$ of two possible
sources and assigns prior probability $\lambda \in [0,1]$ to source $P$.
Before learning which source generates the symbol, the encoder uses the 
predictive distribution
\begin{align}
\label{eq: pmf R_lambda}
\widetilde{R}_\lambda:=\lambda P+(1-\lambda)Q, \qquad \lambda \in [0,1].
\end{align}
With the notation of \cite{NishiyamaS20} and Lemma~\ref{lemma: UB by chi squares}, 
we have $\widetilde{R}_\lambda \equiv R_{1-\lambda}$ for every $\lambda \in [0,1]$. 
For convenience, we use the parametrization $\widetilde{R}_\lambda$ throughout 
this section.

\smallskip 
When the generating law is $P$, how much shorter is the ideal codelength
on average if the encoder uses $\widetilde{R}_\lambda$ instead of $Q$ alone?

For $D(P\|Q)<\infty$, we call the difference
\begin{align}
\label{eq: redundancy reduction by mixing}
  \Delta_\lambda(P,Q)
  := D(P\|Q)-D(P\|\widetilde{R}_\lambda)
  = \mathbb{E}_P\left[\log\frac{\widetilde{R}_\lambda(X)}{Q(X)}\right]
\end{align}
the \emph{redundancy reduction by mixing}. With base-two logarithms,
$\Delta_\lambda(P,Q)$ is the expected saving in bits per symbol when 
\(P\) generates the data, or equivalently, the expected reduction in 
logarithmic loss under $P$~\cite{GneitingRaftery2007}.
The residual redundancy $D(P\|\widetilde{R}_\lambda)$ is the skew divergence with
parameter $1-\lambda$ in the convention of~\cite{Lee1999}.
Convexity of relative entropy gives
\begin{align}
\label{eq: LB by convexity of KL}
  \Delta_\lambda(P,Q) \geq \lambda D(P\|Q), \qquad \lambda \in [0,1],
\end{align}
so assigning weight $\lambda$ to the true source recovers at least
that fraction of the original redundancy.

\begin{proposition}
\label{proposition: bounds on Delta}
Let \(P\) and \(Q\) be distinct probability mass functions on a finite or
countably infinite alphabet \(\mathcal X\) such that
\(A := \chi^2(P\|Q) < \infty\) and \(B := \chi^2(Q\|P) < \infty\).
For \( \lambda \in [0,1] \) and \( n \in \mathbb{N} \), define
\begin{align}
\label{eq: a,b_lambda}
  & a_\lambda^{(n)}:=\lambda^2 \alpha_n,\qquad \qquad \quad 
  b_\lambda^{(n)}:=\frac{\lambda^2 \alpha_n \beta_n}{\lambda \alpha_n + (1-\lambda) \beta_n + \lambda (1-\lambda) \alpha_n \beta_n}, \\
  \label{eq: alpha, beta}
  & \alpha_n:=(1+A)^n-1, \qquad \beta_n := (1+B)^n-1,
\end{align}
let the function \( F \colon (0, \infty)^2 \to (0, \infty) \) be as given in \eqref{eq: function F},
and let \( P^n \) and \( Q^n \) be the product probability mass functions given by 
\begin{align}
\label{eq: product measures}
P^n(x^n) = \prod_{i=1}^n P(x_i), \qquad Q^n(x^n) = \prod_{i=1}^n Q(x_i), \qquad x^n := (x_1, \ldots, x_n) \in \mathcal{X}^n.
\end{align} 
Then, for every \( \lambda \in [0,1] \) and \( n \in \mathbb{N} \),
\begin{align}
\label{eq: LB Delta}
\log\Bigl( 1 + \lambda \bigl[ \exp \bigl(n D(P\|Q) \bigr) - 1 \bigr] \Bigr) & \leq   
\Delta_\lambda(P^n,Q^n) \\
\label{eq: UB Delta}
&\leq \frac{F\bigl(a_\lambda^{(n)}, b_\lambda^{(n)}\bigr) + (1-\lambda) F\bigl(b_\lambda^{(n)}, a_\lambda^{(n)}\bigr)}{\lambda}.
\end{align}
Furthermore, the upper bound on \( \Delta_\lambda(P^n,Q^n) \) is locally tight in the sense that, for all \( n \in \mathbb{N} \), 
\begin{align}
\label{eq1: local tightness}
\lim_{\lambda \to 0^+} \frac{\Delta_\lambda(P^n,Q^n)}{\lambda} &= \bigl[ (1+A)^n - 1 \bigr] \, \log e \\
\label{eq2: local tightness}
&= \lim_{\lambda \to 0^+} \frac{F\bigl(a_\lambda^{(n)}, b_\lambda^{(n)}\bigr) + (1-\lambda) F\bigl(b_\lambda^{(n)}, a_\lambda^{(n)}\bigr)}{\lambda^2}.
\end{align}
\end{proposition}

Before proving Proposition~\ref{proposition: bounds on Delta}, we discuss the usefulness of the 
lower and upper bounds on \( \Delta_\lambda(P^n,Q^n) \), as given in \eqref{eq: LB Delta} and 
\eqref{eq: UB Delta}, respectively. 

\begin{remark}[Computational complexity of \( \Delta_\lambda(P^n,Q^n) \) and its bounds]
{\em 
Suppose that the set $\mathcal{X}$ is finite. Although 
\( P^n \) and \( Q^n \) are product measures, for \(P\ne Q\) and \(n\ge 2\), the mixture
\begin{align}
\label{eq: pmf R_lambda^n}
\widetilde{R}_\lambda^{(n)} := \lambda P^n+(1-\lambda)Q^n, \qquad \lambda \in [0,1],
\end{align}
is a nonproduct measure for every \(\lambda\in(0,1)\). Consequently, the
relative entropy \(D(P^n\|\widetilde{R}_\lambda^{(n)})\) does not tensorize. The defining 
sum for \(\Delta_\lambda(P^n,Q^n)\) has \(\lvert\mathcal{X}\rvert^n\) terms, 
a number that grows exponentially with \(n\). Grouping sequences of the same 
type reduces the sum to at most \(\binom{n+k-1}{k-1}\) terms, where
\(k:=\lvert\mathcal{X}\rvert\). For fixed \(k\), the number of types is 
asymptotic to \(\frac{n^{k-1}}{(k-1)!}\) as \(n\to\infty\). Its polynomial degree
\(k-1=\lvert\mathcal{X}\rvert-1\) increases linearly with the alphabet size.

By contrast, the upper bound on \(\Delta_\lambda(P^n,Q^n)\) in
\eqref{eq: UB Delta} depends on \(P\) and \(Q\) only through the
chi-squared divergences \(\chi^2(P\|Q)\) and \(\chi^2(Q\|P)\)
(see \eqref{eq: function F}, \eqref{eq: a,b_lambda}, and \eqref{eq: alpha, beta}). 
Evaluating this bound therefore requires two sums of \(\lvert\mathcal{X}\rvert\)
terms each. Similarly, the lower bound on the left-hand side of
\eqref{eq: LB Delta} depends on \(P\) and \(Q\) only through the
relative entropy \(D(P\|Q)\), which requires a single sum of
\(\lvert\mathcal{X}\rvert\) terms.}
\end{remark}

We next prove Proposition~\ref{proposition: bounds on Delta}.
\begin{proof}
It suffices to prove Proposition~\ref{proposition: bounds on Delta} for \(n=1\).
Let \(A := \chi^2(P\|Q)\) and \(B := \chi^2(Q\|P)\), both finite by assumption.
By the tensorization identities for relative entropy and the chi-squared divergence,
\begin{align}
\label{eq: tensorization of KL and chi^2}
D(P^n \| Q^n) &= n D(P \| Q), &
\chi^2(P^n \| Q^n) &= (1+A)^n - 1, &
\chi^2(Q^n \| P^n) &= (1+B)^n - 1.
\end{align}
All three divergences are finite, since \(D(P\|Q) \leq \log(1+A)\) (see \eqref{eq: well-known}).
To prove the result for \( n=1 \), we use the following consequence of
the weighted Pythagorean identity for relative entropy. 
\begin{lemma}[Generalized Pythagorean identity]
\label{lemma: Generalized Pythagorean identity}
Let \(P\), \(Q\), and \(W\) be probability mass functions on a common
finite or countably infinite alphabet \(\mathcal{X}\), and suppose that
\(D(P\|Q)\), \(D(P\|W)\), and \(D(Q\|W)\) are finite.
For \(\lambda \in (0,1]\), let \(\widetilde{R}_\lambda\) be the convex combination
in \eqref{eq: pmf R_lambda}. Then
\begin{align}
\label{eq: Generalized Pythagorean identity}
\Delta_\lambda(P,Q)
= D(P\|W) - D(P\|\widetilde{R}_\lambda)
   + \frac{1-\lambda}{\lambda}\,D(Q\|W) + \frac{1}{\lambda}
   \bigl[D(\widetilde{R}_\lambda\|Q) - D(\widetilde{R}_\lambda\|W)\bigr].
\end{align}
\end{lemma}

\begin{proof}
By \eqref{eq: redundancy reduction by mixing},
\begin{align*}
\Delta_\lambda(P,Q)
&= D(P\|Q) - D(P\|\widetilde{R}_\lambda) \\
&= \sum_{x \in \mathcal{X}} P(x) \log\frac{P(x)}{Q(x)}
   - \sum_{x \in \mathcal{X}} P(x) \log\frac{P(x)}{\widetilde{R}_\lambda(x)} \\
&= \sum_{x \in \mathcal{X}} P(x) \log\frac{\widetilde{R}_\lambda(x)}{W(x)}
   + \sum_{x \in \mathcal{X}} P(x) \log\frac{W(x)}{Q(x)}.
\end{align*}
The first sum satisfies
\begin{align*}
\sum_{x \in \mathcal{X}} P(x) \log\frac{\widetilde{R}_\lambda(x)}{W(x)}
&= \sum_{x \in \mathcal{X}} P(x) \log\frac{P(x)}{W(x)}
   - \sum_{x \in \mathcal{X}} P(x) \log\frac{P(x)}{\widetilde{R}_\lambda(x)} \\
&= D(P\|W) - D(P\|\widetilde{R}_\lambda).
\end{align*}
For the second sum, substituting
\(\lambda P(x)=\widetilde{R}_\lambda(x)-(1-\lambda)Q(x)\) gives
\begin{align*}
\sum_{x \in \mathcal{X}} P(x) \log\frac{W(x)}{Q(x)}
&= \frac{1}{\lambda}
   \sum_{x \in \mathcal{X}}
   \bigl[\widetilde{R}_\lambda(x)-(1-\lambda)Q(x)\bigr]
   \log\frac{W(x)}{Q(x)} \\
&= \frac{1}{\lambda}
   \sum_{x \in \mathcal{X}} \widetilde{R}_\lambda(x)
   \left[\log\frac{\widetilde{R}_\lambda(x)}{Q(x)}
        -\log\frac{\widetilde{R}_\lambda(x)}{W(x)}\right]
   + \frac{1-\lambda}{\lambda}
   \sum_{x \in \mathcal{X}} Q(x) \log\frac{Q(x)}{W(x)} \\
&= \frac{1}{\lambda}
   \bigl[D(\widetilde{R}_\lambda\|Q)-D(\widetilde{R}_\lambda\|W)\bigr]
   + \frac{1-\lambda}{\lambda}\,D(Q\|W).
\end{align*}
Adding the two expressions yields \eqref{eq: Generalized Pythagorean identity}.
\end{proof}

The Pythagorean identity for the relative entropy follows as a special 
case of the identity in \eqref{eq: Generalized Pythagorean identity}.
\begin{corollary}[Pythagorean identity for the relative entropy]
Let \(P\), \(Q\), and \(W\) be probability mass functions on a common
finite or countably infinite alphabet, and suppose that \(D(P\|Q)\),
\(D(P\|W)\), and \(D(Q\|W)\) are finite. Let \(M:=\tfrac12(P+Q)\).
Then
\begin{align}
\label{eq: Pythagorean identity}
D(P\|W)+D(Q\|W)
= D(P\|M)+D(Q\|M)+2D(M\|W).
\end{align}
\end{corollary}

\begin{proof}
Taking \(\lambda=\tfrac12\) in
\eqref{eq: Generalized Pythagorean identity} and using \(\widetilde{R}_{1/2}=M\)
gives
\begin{align*}
\Delta_{1/2}(P,Q)
&= D(P\|W)-D(P\|M)+D(Q\|W)
   +2D(M\|Q)-2D(M\|W).
\end{align*}
Taking \(W=M\) in the last equality gives
\[
\Delta_{1/2}(P,Q)=D(Q\|M)+2D(M\|Q).
\]
Equating the two expressions for \(\Delta_{1/2}(P,Q)\) and canceling
\(2D(M\|Q)\) gives \eqref{eq: Pythagorean identity} after rearrangement.
\end{proof}

\begin{corollary}
\label{corollary: Pythagorean identity}
Let \(P\) and \(Q\) be probability mass functions on a common finite or
countably infinite alphabet, and suppose that \(D(P\|Q)<\infty\).
For \(\lambda\in(0,1]\), let \(\widetilde{R}_\lambda=\lambda P+(1-\lambda)Q\).
Then
\begin{align}
\label{eq: Delta in terms of mixture divergences}
\Delta_\lambda(P,Q)
= \frac{1}{\lambda}\,D(\widetilde{R}_\lambda\|Q)
  + \frac{1-\lambda}{\lambda}\,D(Q\|\widetilde{R}_\lambda),
\end{align}
where the second term is omitted at \(\lambda=1\).
\end{corollary}
\begin{proof}
Eq.~\eqref{eq: Delta in terms of mixture divergences} follows from \eqref{eq: Generalized Pythagorean identity} 
as a special case by setting \( W := \widetilde{R}_\lambda \).
\end{proof}

Combining Proposition~\ref{proposition: KL and chi^2} and Corollary~\ref{corollary: Pythagorean identity}, we get 
for all \(\lambda\in(0,1]\)
\begin{align}
\Delta_\lambda(P,Q)
&= \frac{1}{\lambda}\; D(\widetilde{R}_\lambda\|Q)
  + \left(\frac{1-\lambda}{\lambda} \right) D(Q\|\widetilde{R}_\lambda) \nonumber \\[0.1cm]
\label{eq1: 22.11.26}  
&\leq \frac{1}{\lambda} \;  F\bigl( \chi^2(\widetilde{R}_\lambda \| Q), \; \chi^2(Q \| \widetilde{R}_{\lambda}) \bigr)
+ \left( \frac{1-\lambda}{\lambda} \right) F\bigl( \chi^2(Q \| \widetilde{R}_{\lambda}), \; \chi^2(\widetilde{R}_\lambda \| Q) \bigr),
\end{align}
where the function $F \colon (0, \infty)^2 \to (0, \infty)$ is given in \eqref{eq: function F}.
Let \( a_\lambda := a_\lambda^{(1)} \) and \( b_\lambda := b_\lambda^{(1)} \) (see \eqref{eq: a,b_lambda} and \eqref{eq: alpha, beta}).
By \eqref{eq: pmf R_lambda} and Lemma~\ref{lemma: UB by chi squares}, 
\begin{align}
\chi^2(Q \| \widetilde{R}_\lambda) &= \sum_{x \in \mathcal{X}} \frac{\bigl( \widetilde{R}_\lambda(x) - Q(x) \bigr)^2}{\widetilde{R}_\lambda(x)} \nonumber  \\
&= \lambda^2 \sum_{x \in \mathcal{X}} \frac{\bigl( P(x) - Q(x) \bigr)^2}{R_{1-\lambda}(x)} \nonumber \\
&\leq \frac{\lambda^2 AB}{\lambda A + (1-\lambda)B + \lambda (1-\lambda) AB} \nonumber \\
\label{eq2: 22.11.26}
&= b_\lambda, 
\end{align}
and 
\begin{align}
\chi^2(\widetilde{R}_\lambda \| Q) &= \sum_{x \in \mathcal{X}} \frac{\bigl( \widetilde{R}_\lambda(x) - Q(x) \bigr)^2}{Q(x)} \nonumber \\
&= \lambda^2 \sum_{x \in \mathcal{X}} \frac{\bigl( P(x) - Q(x) \bigr)^2}{Q(x)} \nonumber \\
&= \lambda^2 \, \chi^2(P \| Q) \nonumber \\
\label{eq3: 22.11.26}
&= a_\lambda. 
\end{align}
Since $F$ is strictly increasing in both arguments, as shown in the proof of Corollary~\ref{corollary1: KL and chi^2}, 
\eqref{eq1: 22.11.26}--\eqref{eq3: 22.11.26} give
\begin{align}
\label{eq4: 22.11.26}
\Delta_\lambda(P,Q)
\leq \frac{1}{\lambda} \;  F(a_\lambda, b_\lambda)
+ \left( \frac{1-\lambda}{\lambda} \right) F(b_\lambda, a_\lambda).
\end{align}
This proves \eqref{eq: UB Delta} for $n=1$. By \cite[Eq.~(51)]{NishiyamaS20} and the relation 
$\widetilde{R}_\lambda = R_{1-\lambda}$ for every $\lambda \in [0,1]$,
\begin{align}
\label{eq5: 22.11.26}
D(P \| \widetilde{R}_\lambda) \leq \log \left( \frac{1}{\lambda + (1-\lambda) \, \exp\bigl(-D(P\|Q)\bigr)} \right),
\end{align} 
which then gives 
\begin{align}
\Delta_\lambda(P,Q) &= D(P \| Q) - D(P \| \widetilde{R}_\lambda) \nonumber \\
&\geq D(P \| Q) + \log \left( \lambda + (1-\lambda) \, \exp\bigl(-D(P\|Q)\bigr) \right) \nonumber \\
&= \log \Bigl( 1 + \lambda \bigl[ \exp \bigl( D(P\|Q) \bigr) - 1 \bigr] \Bigr).
\end{align}
This proves the lower bound on the left-hand side of \eqref{eq: LB Delta} for $n=1$.

\medskip
We next prove \eqref{eq1: local tightness} and \eqref{eq2: local tightness}
for \( n=1 \). On the common support of \( P \) and \( Q \), let
\( L(x):=P(x)/Q(x) \). For \( 0<\lambda\leq\tfrac12 \), the mean-value
theorem gives
\[
  \frac{1}{\lambda}
  \left| \log \frac{\widetilde{R}_\lambda(x)}{Q(x)} \right|
  = \frac{1}{\lambda}
    \left| \log \bigl(1+\lambda(L(x)-1)\bigr) \right|
  \leq 2\,|L(x)-1|\,\log e.
\]
Since \( \mathbb{E}_P[L(X)]=1+A<\infty \), the right-hand side is
integrable under \( P \). Dominated convergence therefore justifies
interchanging the limit and the sum in the following calculation.
\begin{align}
\lim_{\lambda \to 0^+} \frac{\Delta_\lambda(P,Q)}{\lambda}
&= \lim_{\lambda \to 0^+} \frac{D(P \| Q) - D(P \| \widetilde{R}_\lambda)}{\lambda} \nonumber \\
&= \lim_{\lambda \to 0^+} \frac1{\lambda} \sum_{x \in \mathcal{X}} P(x) \log \left( \frac{\widetilde{R}_\lambda(x)}{Q(x)} \right) \nonumber \\
&= \sum_{x \in \mathcal{X}} \left\{ P(x) \, \lim_{\lambda \to 0^+} \frac1{\lambda} \log \left( \frac{\widetilde{R}_\lambda(x)}{Q(x)} \right) \right\} \nonumber \\
&= \sum_{x \in \mathcal{X}} \left\{ P(x) \, \left.\frac{\partial}{\partial\lambda} \bigl\{ \log \widetilde{R}_\lambda(x)\bigr\} \right|_{\lambda=0^+} \right\} \nonumber \\
&= \sum_{x \in \mathcal{X}} \left\{ P(x) \, \lim_{\lambda \to 0^+} \frac{P(x)-Q(x)}{\lambda P(x) + (1-\lambda) Q(x)} \right\} \, \log e \nonumber \\
&= \sum_{x \in \mathcal{X}} \left\{ \frac{P(x) \, \bigl( P(x) - Q(x) \bigr)}{Q(x)} \right\} \log e \nonumber \\
&= \sum_{x \in \mathcal{X}} \frac{\bigl( P(x) - Q(x) \bigr)^2}{Q(x)} \; \log e \nonumber \\
&= \chi^2(P \| Q) \, \log e.
\end{align}
This proves \eqref{eq1: local tightness} for \( n=1 \). To prove \eqref{eq2: local tightness} for \( n=1 \), we state and prove the following lemma. 

\begin{lemma}
\label{lem:limits-F}
Let $F$ be given by \eqref{eq: function F}, and let $A,B>0$ be finite.
For $0<\lambda<1$, set
\begin{align}
\label{eq: a,b_lambda; n=1}
  a_\lambda := \lambda^2 A,\qquad
  b_\lambda := \frac{\lambda^2 AB}
  {\lambda A+(1-\lambda)B+\lambda(1-\lambda)AB}.
\end{align}
Then
\begin{align}
\label{eq:limits-F}
  \lim_{\lambda \to 0^+}\frac{F(a_\lambda,b_\lambda)}{\lambda^2}
  =\lim_{\lambda \to 0^+}\frac{F(b_\lambda,a_\lambda)}{\lambda^2}
  =\tfrac12 A \log e.
\end{align}
\end{lemma}

\begin{proof}
Write $b_\lambda=\lambda^2 c_\lambda$, where
\[
  c_\lambda :=
  \frac{AB}{\lambda A+(1-\lambda)B+\lambda(1-\lambda)AB} \; \xrightarrow[\lambda \to 0^+]{} \; A.
\]
For either $(u_\lambda,v_\lambda)=(A,c_\lambda)$ or $(u_\lambda,v_\lambda)=(c_\lambda,A)$,
the integral representation in Lemma~\ref{lem:integral} gives
\[
  \frac{F(\lambda^2 u_\lambda,\lambda^2 v_\lambda)}{\lambda^2}
  =(\log e)\int_0^1
  \frac{tu_\lambda v_\lambda}{(1-t)u_\lambda+tv_\lambda+\lambda^2t(1-t)u_\lambda v_\lambda}\,\mathrm{d}t.
\]
In both cases, $u_\lambda, v_\lambda \to A$ as we let $\lambda \to 0^+$. 
Since $(1-t)u_\lambda+tv_\lambda$ lies between $u_\lambda$ and $v_\lambda$
and $0\le t(1-t)\le\tfrac14$, the denominator converges uniformly
to $A$ on $[0,1]$. The numerator converges uniformly to $tA^2$.
As $A>0$, the integrand therefore converges uniformly to $tA$.
By the uniform convergence theorem for integrals, we may pass the
limit through the integral in both cases, obtaining
\[
  (\log e)\int_0^1 tA\,\mathrm{d}t=\tfrac12 A\log e.
\]
\end{proof}

For $n=1$, we show that the ratio of the upper bound on
$\Delta_\lambda(P,Q)$ in \eqref{eq: UB Delta} to $\lambda$
also converges, as $\lambda \to 0^+$, to \( \chi^2(P \| Q) \, \log e \).
Indeed, by Lemma~\ref{lem:limits-F},
\begin{align*}
\lim_{\lambda \to 0^+} \frac{F\bigl(a_\lambda, b_\lambda \bigr) 
+ (1-\lambda) F\bigl(b_\lambda, a_\lambda \bigr)}{\lambda^2}
&= \lim_{\lambda \to 0^+} \frac{F\bigl(a_\lambda, b_\lambda \bigr)}{\lambda^2} 
+ \lim_{\lambda \to 0^+} \frac{F\bigl(b_\lambda, a_\lambda \bigr)}{\lambda^2} \\
&= A \log e \\
&= \chi^2(P \| Q) \, \log e.
\end{align*} 
Hence, the upper bound on \( \Delta_\lambda(P,Q) \) is locally tight at \( \lambda = 0 \). 
This completes the proof of Proposition~\ref{proposition: bounds on Delta} for $n=1$.
As explained in the first paragraph of this proof, the result extends directly to
$n \geq 2$ by the tensorization properties of the relative entropy and ordered 
chi-squared divergences (see \eqref{eq: tensorization of KL and chi^2}).
\end{proof}

\begin{remark}
{\em In contrast to the upper bound on $\Delta_\lambda(P,Q)$, the ratio of its lower bound 
to $\lambda$ does not converge, as $\lambda \to 0^+$, to \( \chi^2(P \| Q) \, \log e \).
Indeed,
\begin{align}
\lim_{\lambda \to 0^+} \frac{\log \Bigl( 1 + \lambda \bigl[ \exp \bigl( D(P\|Q) \bigr) - 1 \bigr] \Bigr)}{\lambda} 
= \bigl[ \exp \bigl( D(P\|Q) \bigr) - 1 \bigr] \, \log e.
\end{align}
Since \(P\ne Q\) and \(P\) and \(Q\) have the same support,
\(P(X)/Q(X)\) is nonconstant under \(P\). Hence, strict Jensen's
inequality gives \(D(P\|Q)<\log\mathbb{E}_P[P(X)/Q(X)]=D_2(P\|Q)\),
so that
\begin{align*}
\exp \bigl( D(P\|Q) \bigr) - 1 &< \exp \bigl( D_2(P\|Q) \bigr) - 1 \\
&= \chi^2(P \| Q).
\end{align*}}
\end{remark}

\subsection*{Funding}
This research received no external funding.

\subsection*{Acknowledgments}
After completing the first draft, the author used ChatGPT~6.1 (OpenAI) 
to recheck selected analytical calculations, help identify potentially 
relevant references, and assist in editing the manuscript for clarity 
and style. The author takes full responsibility for the content of the 
manuscript.

\subsection*{Conflicts of interest}
The author declares no conflicts of interest.

\appendix

\section{An inequality for the monotonicity argument in
Proposition~\ref{prop:binomial-prescribed-divergences}}
\label{app:binary-relative-entropy}

\setcounter{equation}{0}
\renewcommand{\theequation}{\thesection\arabic{equation}}

\setcounter{lemma}{0}
\renewcommand*{\thelemma}{\thesection\arabic{lemma}}

\providecommand*{\theHlemma}{}
\renewcommand*{\theHlemma}{appendix.\thesection.\arabic{lemma}}

We use the notation introduced in the proof of
Proposition~\ref{prop:binomial-prescribed-divergences}.
The following lemma gives the differential inequality for $M(a,b)$
used there to show that $J'(t)>0$ for every $t>0$.

\begin{lemma}[A differential inequality for binary relative entropy]
\label{lem:binary-relative-entropy-differential-inequality}
Let $p$ and $q$ be differentiable functions of $a,b>0$ satisfying
\[
0<q<p<1,\qquad (p-q)^2=a\,q(1-q)=b\,p(1-p),
\]
and let $M(a,b):=d(p\|q)$. Then
\begin{equation}\label{eq:radial-inequality}
(1+a)\ln(1+a)\,M_a(a,b)
+(1+b)\ln(1+b)\,M_b(a,b)>M(a,b).
\end{equation}
\end{lemma}

\begin{proof}
Define the two likelihood ratios
\begin{equation}\label{eq: likelihoods r,s}
r:=\frac{p}{q},\qquad s:=\frac{1-p}{1-q}.
\end{equation}
Since $p>q$, we have $0<s<1<r$. Solving $p=rq$ and
$1-p=s(1-q)$ gives
\[
q=\frac{1-s}{r-s},\qquad p=\frac{r(1-s)}{r-s}.
\]
Moreover, $r-1=(p-q)/q$ and $1-s=(p-q)/(1-q)$, so
\[
a=(r-1)(1-s),\qquad b=\frac{(r-1)(1-s)}{rs}.
\]

The identity $rs=a/b$ suggests taking the geometric mean
$v:=\sqrt{rs}$ as one parameter. Since $\ln v$ is the midpoint
of $\ln r$ and $\ln s$, it is natural to take their half-difference
$z:=\tfrac12\ln(r/s)$ as the second parameter. Then
\[
\ln r=\ln v+z,\qquad \ln s=\ln v-z,
\]
so the weighted average $p\ln r+(1-p)\ln s$ reduces to
$\ln v+z(2p-1)$. Exponentiating gives $r=ve^z$ and $s=ve^{-z}$,
so the condition $0<s<1<r$ is equivalent to $z>0$ and
$e^{-z}<v<e^z$. The sum and difference of the likelihood ratios
are $r+s=2v\cosh z$ and $r-s=2v\sinh z$. Writing $C:=\cosh z$
and $S:=\sinh z$, and using $rs=v^2$, we obtain
\begin{equation}\label{eq:binary-hyperbolic-parameters}
1+a=v(2C-v),\qquad 1+b=\frac{2Cv-1}{v^2},\qquad
2p-1=\frac{r+s-2rs}{r-s}=\frac{C-v}{S}.
\end{equation}
By the definitions of $M$, $r$, and $s$, we have
\begin{align}
\frac{M(a,b)}{\log e}
&=p\ln r+(1-p)\ln s \nonumber\\
&=\ln v+z(2p-1) \nonumber\\
&=\ln v+\frac{z(C-v)}{S},
\label{eq:binary-hyperbolic-entropy}
\end{align}
where the last equality follows from
\eqref{eq:binary-hyperbolic-parameters}.

To compute $M_a$ and $M_b$, denote the final expression in
\eqref{eq:binary-hyperbolic-entropy} by $\Phi(v,z)$.
Differentiating \eqref{eq:binary-hyperbolic-parameters} and the
expression for $\Phi$, and using $C^2-S^2=1$ to simplify $\Phi_z$,
we obtain
\begin{subequations}\label{eq: partial derivatives a,b,Phi}
\begin{align}
a_v&=2(C-v),                 & a_z&=2vS,\\
b_v&=\frac{2(1-Cv)}{v^3},    & b_z&=\frac{2S}{v},\\
\Phi_v&=\frac{1}{v}-\frac{z}{S}, &
\Phi_z&=\frac{C-v}{S}+\frac{z(Cv-1)}{S^2}.
\end{align}
\end{subequations}
The Jacobian determinant of $(a,b)$ with respect to $(v,z)$ is
\[
a_vb_z-a_zb_v
=\frac{4S}{v^2}(2Cv-v^2-1)
=\frac{4Sa}{v^2}>0.
\]
Since $\Phi(v,z)=M(a(v,z),b(v,z))/(\log e)$, the chain rule gives
\begin{equation}\label{eq: identities 3}
\Phi_v=a_v\frac{M_a}{\log e}+b_v\frac{M_b}{\log e},
\qquad
\Phi_z=a_z\frac{M_a}{\log e}+b_z\frac{M_b}{\log e},
\end{equation}
where $M_a$ and $M_b$ are evaluated at $(a(v,z),b(v,z))$.
The determinant of this system is $a_vb_z-a_zb_v>0$, so it
uniquely determines $M_a/(\log e)$ and $M_b/(\log e)$.

To express the solution, set $U:=CS-z$ and $V:=zC-S$.
Since $C^2-S^2=1$, we have
\[
CU+V=S^3,\qquad U+CV=zS^2.
\]
Substituting the expressions for $a_v$ and $b_v$ from
\eqref{eq: partial derivatives a,b,Phi} and using these two
identities gives
\begin{subequations}
\begin{align}
a_v\frac{U}{2vS^3}+b_v\frac{v^2V}{2S^3}
&=\frac{(C-v)U+(1-Cv)V}{vS^3} \nonumber\\
&=\frac{(CU+V)-v(U+CV)}{vS^3}
 =\frac{1}{v}-\frac{z}{S}=\Phi_v.
\label{eq: identity1}
\end{align}
For the derivative with respect to $z$, the definitions of $U$ and
$V$ give $U+vV=S(C-v)+z(Cv-1)$. Substituting the expressions
for $a_z$ and $b_z$ therefore yields
\begin{equation}\label{eq: identity2}
a_z\frac{U}{2vS^3}+b_z\frac{v^2V}{2S^3}
=\frac{U+vV}{S^2}
=\frac{C-v}{S}+\frac{z(Cv-1)}{S^2}=\Phi_z.
\end{equation}
\end{subequations}
By uniqueness of the solution to \eqref{eq: identities 3},
\eqref{eq: identity1} and \eqref{eq: identity2} give
\begin{equation}\label{eq: identities 4}
\frac{M_a}{\log e}=\frac{U}{2vS^3},
\qquad
\frac{M_b}{\log e}=\frac{v^2V}{2S^3}.
\end{equation}

Fix $z>0$ and write $a=a(v,z)$ and $b=b(v,z)$. Define
\begin{equation}\label{eq: G_z}
G_z(v):=\frac{2S^3}{\log e}
\Bigl[(1+a)\ln(1+a)\,M_a(a,b)
     +(1+b)\ln(1+b)\,M_b(a,b)-M(a,b)\Bigr].
\end{equation}
Since the factor $2S^3/(\log e)$ is positive, it suffices to show
that $G_z(v)>0$ for $e^{-z}<v<e^z$. By
\eqref{eq: identities 4} and \eqref{eq:binary-hyperbolic-parameters},
\[
\frac{2S^3}{\log e}(1+a)M_a=U(2C-v),\qquad
\frac{2S^3}{\log e}(1+b)M_b=V(2Cv-1).
\]
Substituting these expressions and \eqref{eq:binary-hyperbolic-entropy}
into \eqref{eq: G_z} gives
\begin{align}
G_z(v)
={}&U(2C-v)\ln\bigl(v(2C-v)\bigr)
   +V(2Cv-1)\ln\frac{2Cv-1}{v^2} \nonumber\\
 &-2S^3\ln v-2zS^2(C-v).
\label{eq2: G_z}
\end{align}
Expanding the logarithms and using $CU+V=S^3$ yields
\begin{align}
G_z(v)
={}&U(2C-v)\ln(2C-v)+V(2Cv-1)\ln(2Cv-1) \nonumber\\
 &-(U+4CV)v\ln v-2zS^2(C-v).
\label{eq3: G_z}
\end{align}

To prove that $G_z$ is positive on $(e^{-z},e^z)$, we first show
that $G_z$ and $G_z'$ vanish at both endpoints. We then bound the
number of interior zeros of $G_z''$.

The expressions in \eqref{eq2: G_z} and \eqref{eq3: G_z} extend
smoothly to $[e^{-z},e^z]$, since $v$, $2C-v$, and $2Cv-1$ are
positive throughout this interval. Write either endpoint as
$w=e^{\varepsilon z}$, where $\varepsilon\in\{-1,1\}$. Then
\[
2C-w=w^{-1},\qquad 2Cw-1=w^2,\qquad C-w=-\varepsilon S.
\]
Thus $w(2C-w)=1$ and $(2Cw-1)/w^2=1$, so the first two logarithms
in \eqref{eq2: G_z} vanish at $v=w$. Since $\ln w=\varepsilon z$,
the remaining terms give
\[
G_z(w)=-2S^3\ln w-2zS^2(C-w)
      =-2\varepsilon zS^3+2\varepsilon zS^3=0.
\]

Differentiating \eqref{eq3: G_z} with respect to $v$ gives
\begin{align*}
G_z'(v)
={}&-U\bigl(\ln(2C-v)+1\bigr)
   +2CV\bigl(\ln(2Cv-1)+1\bigr)\\
 &-(U+4CV)(\ln v+1)+2zS^2.
\end{align*}
At $v=w$, we have $\ln(2C-w)=-\ln w$ and
$\ln(2Cw-1)=2\ln w$. Collecting terms yields
\[
G_z'(w)
=\bigl[U+4CV-(U+4CV)\bigr]\ln w-2(U+CV)+2zS^2=0,
\]
where the last equality uses $U+CV=zS^2$. Hence
\begin{equation}
\label{eq:Gz-endpoint-conditions}
G_z(e^{-z})=G_z(e^z)=G_z'(e^{-z})=G_z'(e^z)=0.
\end{equation}

Differentiating once more gives
\begin{align}
G_z''(v)
&=\frac{U}{2C-v}+\frac{4C^2V}{2Cv-1}
  -\frac{U+4CV}{v} \nonumber\\
&=\frac{P_z(v)}{v(2C-v)(2Cv-1)},
\label{eq:Gz-second-derivative}
\end{align}
where
\[
P_z(v):=Uv(2Cv-1)+4C^2Vv(2C-v)
        -(U+4CV)(2C-v)(2Cv-1).
\]
The denominator in \eqref{eq:Gz-second-derivative} is positive on
$[e^{-z},e^z]$, and $P_z$ is a quadratic polynomial with leading
coefficient
\[
2CU-4C^2V+2C(U+4CV)=4C(U+CV)=4CzS^2>0.
\]

By \eqref{eq:Gz-endpoint-conditions}, $G_z$ vanishes at both endpoints,
so Rolle's theorem gives a point $c\in(e^{-z},e^z)$ with $G_z'(c)=0$.
Since $G_z'$ also vanishes at both endpoints, applying Rolle's theorem
to $G_z'$ on $[e^{-z},c]$ and $[c,e^z]$ gives two distinct interior
zeros of $G_z''$. By \eqref{eq:Gz-second-derivative}, these are exactly
the two roots of the quadratic polynomial $P_z$. 
Let $\alpha$ denote the smaller root. Since $P_z$ has positive
leading coefficient and the denominator in
\eqref{eq:Gz-second-derivative} is positive, we have
$G_z''>0$ on $(e^{-z},\alpha)$. On this interval, $G_z'$ is
therefore strictly increasing from its endpoint value
$G_z'(e^{-z})=0$. Thus $G_z'>0$ there, and $G_z(e^{-z})=0$
then gives $G_z>0$ throughout the interval $(e^{-z},\alpha)$.

If $G_z$ had an interior zero, Rolle's theorem would give a zero
of $G_z'$ strictly between that zero and each endpoint. Together
with the two endpoint zeros in \eqref{eq:Gz-endpoint-conditions},
these would give four distinct zeros of $G_z'$ and hence at least
three distinct interior zeros of $G_z''$, a contradiction.
Therefore $G_z$ has no interior zero. By continuity, its positivity
on $(e^{-z},\alpha)$ extends to all of $(e^{-z},e^z)$, proving
\eqref{eq:radial-inequality}.
\end{proof}

\end{document}